\documentclass[prl,aps,10pt,longbibliography,superscriptaddress,twocolumn]{revtex4-2}
\usepackage{amsmath}
\usepackage{amssymb}
\usepackage{amsthm}
\usepackage{amsfonts}
\usepackage{enumerate}
\usepackage{latexsym}
\usepackage{bm}
\usepackage{graphicx}
\usepackage{subfigure}
\usepackage{color}
\usepackage[dvipsnames]{xcolor}
\usepackage[normalem]{ulem}
\usepackage[colorlinks]{hyperref}
\hypersetup{
  colorlinks,
  citecolor=Violet,
  linkcolor=Red,
  urlcolor=Blue}
\usepackage{cancel}
\usepackage{mathrsfs}
\usepackage{braket}

\newcommand{\beq}{\begin{equation}}
\newcommand{\eneq}{\end{equation}}
\input{epsf}

\begin{document}

\title{Synchronizing Spectral and Interference Criticalities at an Exceptional Point}

\author{Zeng-Zhao Li}
\email{lizengzhao@iqasz.cn}
\affiliation{International Quantum Academy, Shenzhen, 518048, China}

\author{Xiao Xue}
\email{xiao.xue@ustc.edu.cn}
\affiliation{International Quantum Academy, Shenzhen, 518048, China}
\affiliation{Hefei National Laboratory, University of Science and Technology of China, Hefei 230088, China}

\begin{abstract}
Exceptional points (EPs) are spectral singularities, but whether they can be used to control a distinct criticality of coherent interference remains largely unexplored. Here we show in nonlocal superconducting transport how spectral and interference criticalities can be controllably synchronized. In a minimal superconducting double-quantum-dot system, loss imbalance selectively drives one Bogoliubov sector through an EP. The spectral EP and the CAR--ECT interference criticality form distinct structures in control space whose relative ordering can be continuously tuned. At their synchronization point, the interference boundaries between crossed Andreev reflection (CAR) and elastic cotunneling (ECT) coalesce at the EP and bifurcate beyond it, creating a finite CAR-dominated window. The resulting interference window opens with a square-root critical law, whose exponent also governs the finite-temperature observability scale. Detuning unfolds the two criticalities, while retuning restores their intersection, establishing the synchronization as controlled rather than generic. The bifurcation is directly observable in nonlocal conductance. Our results establish exceptional points as spectral control points for independently defined interference criticalities, opening a route to non-Hermitian control of critical quantum interference.
\end{abstract}

\date{\today}
\maketitle

{\it Introduction}---Interference between coherent pathways is a fundamental mechanism by which microscopic quantum amplitudes are converted into observable responses. A natural question is whether such interference can itself acquire a controllable critical structure, and whether that criticality can be tied to an independently tunable spectral singularity. 
Nonlocal superconducting transport provides a particularly transparent realization of this problem. Elastic cotunneling (ECT) and crossed Andreev reflection (CAR) constitute two competing coherent transfer processes whose balance controls nonlocal charge transport and Cooper-pair splitting~\cite{FalciHekking01epl,RecherLoss01prb,RussoMorpurgo05prl,Yeyati07nphys,HofstetterSchonenberger09nature,MetalidisSchon10prb,MortenBelzig06prb}. 
Their coherent recombination through underlying Bogoliubov propagation channels provides a natural interference structure that can be reorganized spectrally.  
Conventional control relies on level structure, tunnel couplings, interactions, or device geometry~\cite{HofstetterSchonenberger09nature,BordinKouwenhovenDvir23prx}, while recent experiments and discussions have highlighted both the reach of nonlocal Andreev processes and the limitations of identifying CAR and ECT from conductance alone~\cite{FengAndo25nphys,TikhonovKhrapai26nphys,FengAndo26nphys}. 

Exceptional points provide a particularly intriguing spectral control point for this problem. At an EP, eigenvalues and eigenvectors coalesce and the effective non-Hermitian Hamiltonian becomes defective~\cite{AshidaGongUeda20,GanainyChristodoulides18nphys,Bergholtz21rmp}. 
These singularities underlie enhanced sensing~\cite{Wiersig14prl,ChenYang17nature,HodaeiKhajavikhan17nature}, unconventional dynamics~\cite{LiWhaley23prl,LiWhaley24prr}, and broader spectral and topological phenomena~\cite{AshidaGongUeda20,GanainyChristodoulides18nphys,Bergholtz21rmp}. 
Their consequences have also been actively explored in superconducting settings, including pairing, Bogoliubov and Andreev spectra, Josephson transport, and topology~\cite{OhnmachtBelzig25prl,SolowFlensberg25prb,CayaoSchaffer22prb,CayaoSchaffer23prb,LiTrauzettel24prb,CayaoSato24prb,Capecelatro25prb,CayaoSato26arxiv,LiTrauzettel25prb,MaSong25prb}. 
Most of these studies, however, exploit anomalous behavior generated by the EP itself. A conceptually different possibility is to use an EP as a spectral control point for another, independently defined critical structure. Whether such distinct criticalities can be deliberately synchronized, and what observable physics emerges at their intersection, remains largely unexplored. In the present setting, the relevant second structure is the interference criticality separating ECT- and CAR-dominated transport; crucially, it is not a spectral property of the EP itself. 
It is instead set by the relative complex response of the two Bogoliubov propagation sectors.

Here we establish, in a minimal superconducting two-site Bogoliubov–de Gennes system, how spectral and interference criticalities can be controllably synchronized.  
At particle-hole symmetry, the Hamiltonian decomposes into two propagation sectors with effective couplings \(J_\pm=t\pm\Delta\), allowing a local loss imbalance to drive one sector through an EP while the other remains off critical. 
Because ECT and CAR arise from coherent recombination of the two sectors, selectively restructuring one sector reorganizes their interference without requiring a new transport channel.   
We show that the spectral EP and the CAR--ECT interference criticality are distinct structures in control space whose relative ordering can be continuously tuned and deliberately synchronized at a controllable intersection. At this intersection, spectral coalescence is converted into an observable critical bifurcation of interference between pre-existing coherent pathways. 
The superconducting setting therefore provides a directly measurable realization of a broader mechanism linking spectral and interference criticalities.

\begin{figure}[t]
\centering
 \includegraphics[width=0.99\columnwidth]{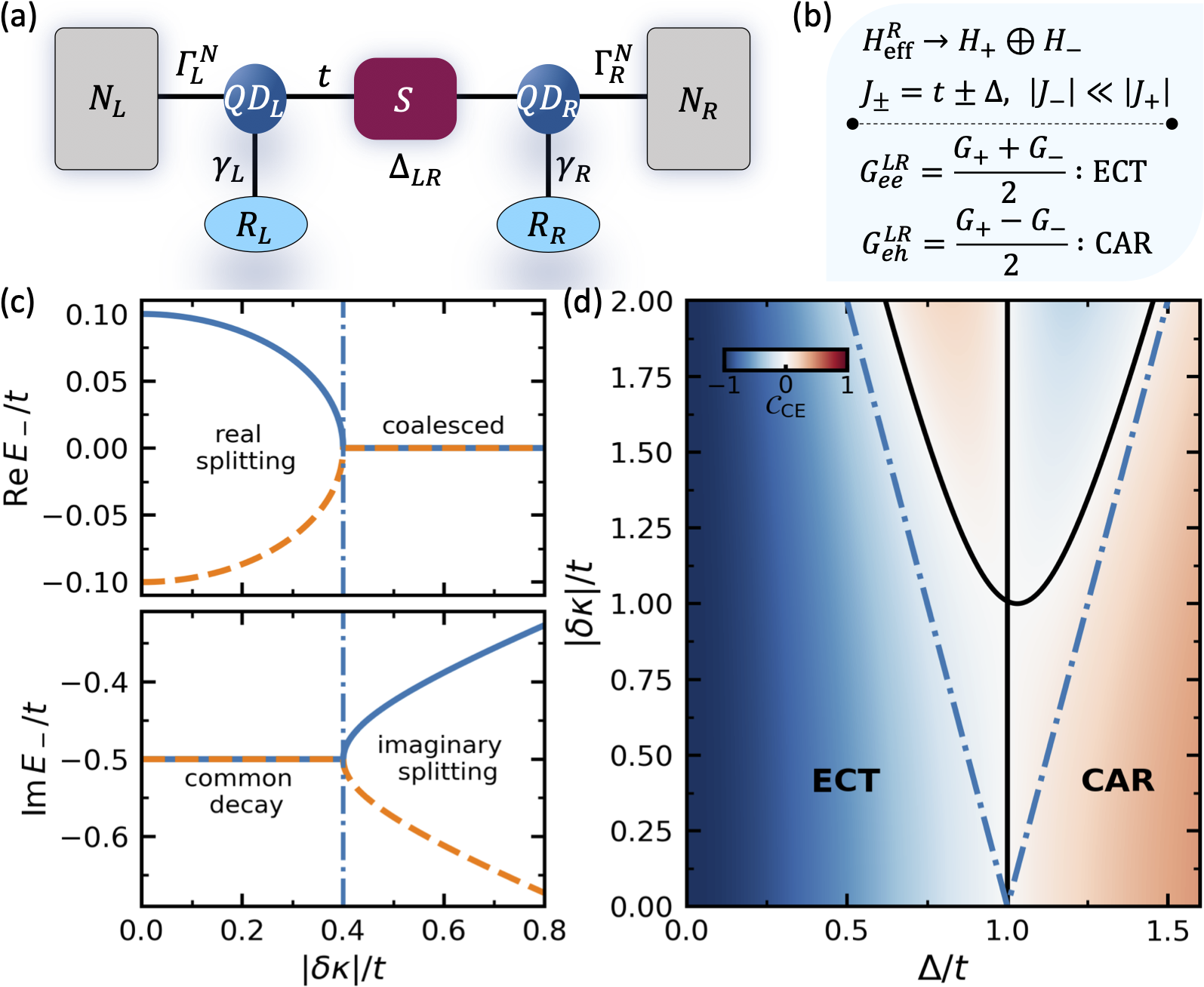} 
\caption{(Color online) {\bf Minimal model, sector-selective exceptional point, and CAR--ECT interference structure.} 
(a) Minimal nonlocal superconducting structure consisting of quantum dots, $L$ and $R$, coupled by normal hopping $t$ and induced nonlocal pairing $\Delta\equiv\Delta_{LR}$, to normal transport leads with rates $\Gamma_L^N$ and $\Gamma_R^N$, and to independently engineered loss reservoirs with rates $\gamma_L$ and $\gamma_R$. 
(b) Bogoliubov-sector decomposition of the effective retarded Hamiltonian. The two sectors have couplings $J_\pm=t\pm\Delta$; for $t\simeq\Delta$, $|J_-|\ll |J_+|$, enabling selective access to the $H_-$ exceptional point. Their coherent recombination gives the ECT and CAR amplitudes through \(G_{ee}^{LR}=(G_++G_-)/2\) and \(G_{eh}^{LR}=(G_+ - G_-)/2\), respectively. 
(c) Spectral transition of the $H_-$ sector for $\Delta=0.9t$: the real parts of the two eigenvalues coalesce at $|\delta\kappa|/t=0.4$, while beyond the exceptional point their imaginary parts split from a common decay rate. 
(d) CAR--ECT contrast $\mathcal{C}_{\rm CE}$ as a function of $\Delta/t$ and $|\delta\kappa|/t$ at fixed $\omega/t=0.5$ and $\bar\kappa/t=1$. The black solid curves mark the CAR–ECT balance $\mathcal{C}_{\rm CE}=0$, while the blue dash-dotted curve denotes the $H_-$ exceptional-point condition $|\delta\kappa|_{\rm EP}^{(-)}=4|t-\Delta|$. Their separation illustrates that the interference boundary is not generically locked to the spectral EP.}
\label{fig:Fig1}
\end{figure}

{\it Model and mechanism}---We consider the minimal nonlocal superconducting structure shown in Fig.~\ref{fig:Fig1}(a), consisting of two quantum dots coupled by normal hopping \(t\) and nonlocal pairing \(\Delta\). Each dot is coupled to a normal transport lead with rate \(\Gamma_\alpha^{N}\) and may additionally be subject to an independently engineered loss channel with rate \(\gamma_\alpha\) (\(\alpha=L,R\)). 
After integrating out these reservoirs in the wide-band limit, the retarded effective Hamiltonian is~\cite{sm}
\begin{equation}
H_{\rm eff}^{R}
=
H_{\rm BdG}
-\frac{i}{2}
\sum_{\alpha=L,R}\kappa_\alpha P_\alpha,
\qquad
\kappa_\alpha=\Gamma_\alpha^{N}+\gamma_\alpha ,
\label{eq:Heff-general}
\end{equation}
where \(P_\alpha\) projects onto the local Nambu subspace of site \(\alpha\).

At particle-hole symmetry, a unitary transformation to a convenient local Nambu basis brings the coherent Hamiltonian to \(H_{\rm BdG} = \tau_x\otimes \left(t\sigma_0+\Delta\sigma_x\right)\), where \(\tau_i\) and \(\sigma_i\) act in site and Nambu spaces, respectively. This representation is unitarily equivalent to the canonical fermionic BdG Hamiltonian; the explicit transformation and particle-hole operator are given in the Supplemental Material~\cite{sm}. Since \(\sigma_x|\pm_{\rm N}\rangle=\pm|\pm_{\rm N}\rangle\), the local Bogoliubov combinations \(|\pm_{\rm N}\rangle=(|e\rangle\pm|h\rangle)/\sqrt2\) block diagonalize \(H_{\rm eff}^{R}\) into two propagation sectors [Fig.~\ref{fig:Fig1}(b)],
\begin{equation}
H_\pm
=
-\frac{i\bar\kappa}{2} I
+J_\pm\tau_x
-i\frac{\delta\kappa}{4}\tau_z,
\qquad
J_\pm=t\pm\Delta ,
\label{eq:Hpm}
\end{equation}
where \(\bar\kappa=(\kappa_L+\kappa_R)/2\) and \(\delta\kappa=\kappa_L-\kappa_R\).

The complex eigenvalues are
\begin{equation}
E_{\pm,\eta}
=
-\frac{i\bar\kappa}{2}
+\eta\sqrt{J_\pm^2-\frac{\delta\kappa^2}{16}},
\qquad \eta=\pm1 ,
\label{eq:spectrum}
\end{equation}
so that each sector possesses an EP at \(|\delta\kappa|_{\rm EP}^{(\pm)}=4|J_\pm|\). 
Pairing separates the EP scales: for \(t\simeq\Delta\), \(|J_-|\ll|J_+|\), allowing \(H_-\) to cross its EP while \(H_+\) remains off critical [Fig.~\ref{fig:Fig1}(b)]. 
The corresponding spectral transition is shown in Fig.~\ref{fig:Fig1}(c), where the real-part splitting collapses at the \(H_-\) EP and is replaced beyond it by an imaginary-part splitting.

The physical electron and hole amplitudes coherently recombine these two sectors. Defining the nonlocal sector propagators as \(G_\pm(\omega)\equiv \langle L|(\omega-H_\pm)^{-1}|R\rangle\), one obtains
\begin{equation}
G_\pm(\omega)=
\frac{J_\pm}
{(\omega+i\kappa_L/2)(\omega+i\kappa_R/2)-J_\pm^2}.
\label{eq:Gpm}
\end{equation}
Transforming back to the electron--hole basis gives \(G_{ee}^{LR}=(G_++G_-)/2\) and \(G_{eh}^{LR}=(G_+-G_-)/2\)
[Fig.~\ref{fig:Fig1}(b)]. 
Consequently, the corresponding nonlocal transmission
probabilities are~\cite{sm}
\begin{equation}
T_{\rm ECT}
=
\Gamma_L^N\Gamma_R^N
\Big|\frac{G_++G_-}{2}\Big|^2, \;
T_{\rm CAR}
=
\Gamma_L^N\Gamma_R^N
\Big|\frac{G_+-G_-}{2}\Big|^2 ,
\label{eq:TECTTCAR}
\end{equation}
and hence
\begin{equation}
T_{\rm CAR}-T_{\rm ECT}
=
-\Gamma_L^N\Gamma_R^N
\,{\rm Re}\!\left(G_+G_-^*\right).
\label{eq:interference}
\end{equation}
For visualization, we define the normalized CAR--ECT contrast \(\mathcal{C}_{\rm CE}=(T_{\rm CAR}-T_{\rm ECT})/(T_{\rm CAR}+T_{\rm ECT})\), with \(\mathcal{C}_{\rm CE}>0(<0)\) denoting CAR- (ECT-) dominated transport and \(\mathcal{C}_{\rm CE}=0\) the balance boundary. 

Equation~(\ref{eq:interference}) exposes the mechanism: CAR and ECT are complementary interference outcomes of the two Bogoliubov propagators.  
Tuning \(H_-\) through its EP changes its complex response relative to the off-critical \(H_+\) sector, thereby reorganizing the CAR--ECT interference and shifting the zeros of \({\rm Re}(G_+G_-^*)\). 
The condition $\operatorname{Re}(G_+G_-^*)=0$ defines the boundary at which CAR and ECT are exactly balanced [Fig.~\ref{fig:Fig1}(d)]. Importantly, this interference boundary is not generically tied to the spectral EP. As we show below, the two critical structures can nevertheless be synchronized, turning the EP into the point at which a CAR-dominated transport window is created.

\begin{figure*}[t]
\centering
\includegraphics[width=1.8\columnwidth]{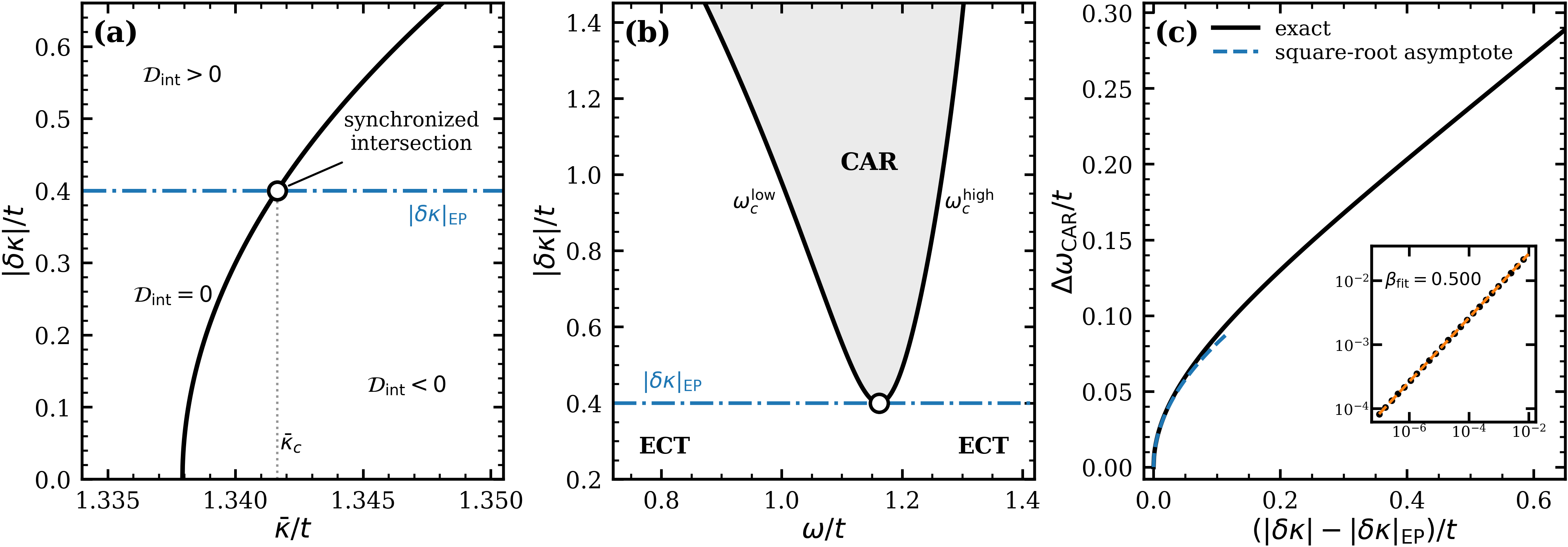} 
\caption{\textbf{Controlled synchronization and exceptional-point-anchored bifurcation of Bogoliubov interference.}
(a) Spectral and interference criticalities in the $(\bar\kappa,|\delta\kappa|)$ control plane for $t=1$ and $\Delta=0.9t$. The blue dash-dotted line marks the $H_-$ exceptional point, $|\delta\kappa|_{\rm EP}=4|J_-|$, while the black curve marks the interference-critical condition $\mathcal D_{\rm int}=0$, separating regions with two real CAR--ECT boundaries ($\mathcal D_{\rm int}>0$) and no real boundaries ($\mathcal D_{\rm int}<0$). Their controlled intersection at $(\bar\kappa_c,|\delta\kappa|_{\rm EP})$ is marked by the open circle; detuning $\bar\kappa$ unfolds the two criticalities and reverses their relative ordering across $\bar\kappa_c$.
(b) At the synchronized cut $\bar\kappa=\bar\kappa_c=\sqrt{2t\Delta}$, the two CAR--ECT balance boundaries $\omega_c^{\rm low}$ and $\omega_c^{\rm high}$ coalesce at the $H_-$ EP and bifurcate above it, opening the shaded CAR-dominated window between ECT-dominated regions. The coalescence occurs at $\omega_*=\sqrt{3t\Delta/2}$.
(c) Critical opening $\Delta\omega_{\rm CAR} =\omega_c^{\rm high}-\omega_c^{\rm low}$. The black curve is the exact result and the blue dashed curve the near-EP square-root asymptote $\Delta\omega_{\rm CAR}\propto (|\delta\kappa|-|\delta\kappa|_{\rm EP})^{1/2}$. The inset shows near-EP points sampled from the exact result and an independent power-law fit, yielding $\beta_{\rm fit}=0.500$.
}
\label{fig:Fig2}
\end{figure*}

The spectral and transport roles of the reservoirs are distinct: the spectral structure depends on the total linewidths \(\kappa_\alpha\), whereas the transport prefactors involve only \(\Gamma_\alpha^N\). The EP can therefore be tuned either through asymmetric transport couplings or through independently controlled auxiliary losses~\cite{AksenovBurmistrov26prb}. 
We first consider the minimal lead-induced realization, \(\gamma_L=\gamma_R=0\), at fixed \(\bar\Gamma^N=(\Gamma_L^N+\Gamma_R^N)/2\), and later show that the same interference mechanism persists when the transport couplings are held fixed and the EP is tuned by auxiliary loss.

{\it Synchronization of spectral and interference criticalities}---An exceptional point is a spectral singularity, whereas the CAR--ECT crossover is governed by interference between the two Bogoliubov sectors. The two critical structures are therefore generically distinct: crossing an EP need not create or eliminate a CAR-dominated transport window. We now show that these two criticalities can nevertheless be deliberately synchronized, converting the EP into a bifurcation point of the transport interference.

From Eq.~(\ref{eq:interference}), the CAR--ECT boundaries are determined by ${\cal F}_{\rm int}(\omega)=0$ where
\({\cal F}_{\rm int}(\omega) = (\omega^2-\Lambda_+)(\omega^2-\Lambda_-) +\bar\kappa^2\omega^2\) with \(\Lambda_s = J_s^2+\frac{\kappa_L\kappa_R}{4}\)~\cite{sm}.
Since ${\cal F}_{\rm int}$ is quadratic in $x=\omega^2$, the existence of real interference boundaries is controlled by its discriminant $\mathcal D_{\rm int}$.
Requiring the two boundaries to coalesce precisely at the $H_-$ exceptional point, $|\delta\kappa|_{\rm EP}=4|J_-|$, selects the critical mean linewidth
\begin{equation}
\bar\kappa_c^2 = \frac{J_+^2-J_-^2}{2} = 2t\Delta .
\label{eq:kappac}
\end{equation}
At this synchronized point,
\begin{equation}
\mathcal D_{\rm int}
=
\frac{J_+^2-J_-^2}{8}
\left(
|\delta\kappa|^2-|\delta\kappa|_{\rm EP}^2
\right),
\label{eq:Dintfactor}
\end{equation}
so that, along the synchronized cut, the sign change of the interference discriminant coincides exactly with crossing the spectral EP.

The synchronization is a controlled intersection rather than an accidental coincidence. Figure~\ref{fig:Fig2}(a) shows the spectral and interference critical structures in the $(\bar\kappa,|\delta\kappa|)$ control plane.
The spectral-EP condition is independent of $\bar\kappa$, whereas the interference-critical condition $\mathcal D_{\rm int}=0$ moves continuously with $\bar\kappa$. 
For $\bar\kappa<\bar\kappa_c$, the interference-critical value of \(|\delta\kappa|\) lies below $|\delta\kappa|_{\rm EP}$; for $\bar\kappa>\bar\kappa_c$, it lies above it.
Thus tuning $\bar\kappa$ continuously reverses their relative ordering through the intersection at $\bar\kappa=\bar\kappa_c$. 
Under weak onsite-energy detuning, sector mixing displaces the intersection, which is restored by retuning $\bar\kappa$; the full unfolding is given in the Supplemental Material~\cite{sm}.

At the synchronized cut, this intersection becomes the interference bifurcation shown in Fig.~\ref{fig:Fig2}(b). Below the EP, $\mathcal D_{\rm int}<0$, so no real CAR--ECT balance boundaries exist and transport is ECT dominated throughout the spectrum.
At the EP, the two boundaries coalesce at \(\omega_*=\sqrt{3t\Delta/2}\). Above it, they split into \(\omega_c^{\rm low}\) and \(\omega_c^{\rm high}\), opening a finite CAR-dominated interval between two ECT-dominated regions.
The synchronized EP therefore marks the onset of an ECT-only $\rightarrow$ ECT--CAR--ECT interference structure, rather than merely shifting a pre-existing crossover.

The window width directly quantifies this bifurcation. Defining
\(
\Delta\omega_{\rm CAR}
\equiv
\omega_c^{\rm high}-\omega_c^{\rm low},
\)
close to the EP,
\begin{equation}
\Delta\omega_{\rm CAR}
\simeq
\sqrt{\frac{2|J_-|}{3}}\,
\bigl(
|\delta\kappa|-|\delta\kappa|_{\rm EP}
\bigr)^{1/2}.
\label{eq:critical-opening}
\end{equation}
As shown in Fig.~\ref{fig:Fig2}(c), the exact window width follows this square-root onset near the bifurcation. 
An independent fit to near-EP points sampled from the exact result using
\(
\Delta\omega_{\rm CAR}/t
=
A_{\rm fit}
[(|\delta\kappa|-|\delta\kappa|_{\rm EP})/t]^{\beta_{\rm fit}}
\)
gives \(\beta_{\rm fit}=0.500\), in agreement with the analytical exponent \(\beta=1/2\).
The exceptional point therefore becomes a controllable source of Bogoliubov-interference criticality: its spectral coalescence is converted into the birth and square-root opening of a CAR-dominated transport window.

\begin{figure}[t]
\centering
\includegraphics[width=0.92\columnwidth]{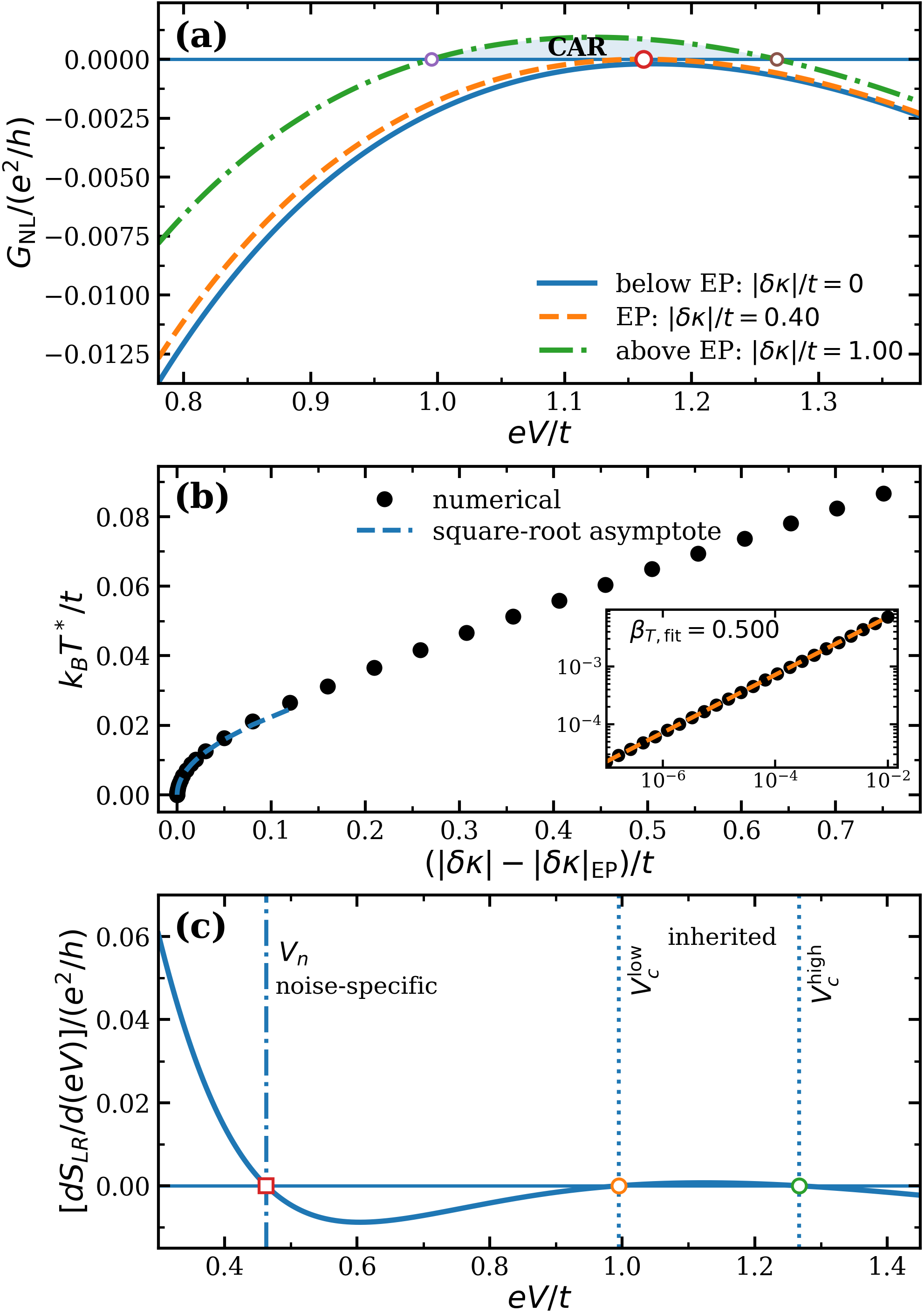} 
\caption{
\textbf{Observable critical bifurcation.}
(a) Zero-temperature nonlocal differential conductance across the synchronized $H_-$ exceptional point, for $|\delta\kappa|/t=0$, $0.40$, and $1.00$ below, at, and above the EP, respectively. Below the EP, $G_{\rm NL}$ remains negative; at the EP it touches zero tangentially; above the EP a finite positive-conductance interval emerges, directly realizing the ECT-only $\rightarrow$ critical-touching $\rightarrow$ ECT--CAR--ECT bifurcation.
(b) Thermal accessibility of the CAR-dominated regime. The symbols show the temperature $T^*$ at which the positive-conductance window closes as a function of distance from the EP. The dashed line is the near-EP square-root asymptote $k_BT^*\propto(|\delta\kappa|-|\delta\kappa|_{\rm EP})^{1/2}$. The inset shows the same near-EP numerical data on logarithmic axes and an independent power-law fit $k_BT^*/t=A_{T,{\rm fit}}[(|\delta\kappa|-|\delta\kappa|_{\rm EP})/t]^{\beta_{T,{\rm fit}}}$, with both fit parameters free, yielding $\beta_{T,{\rm fit}}=0.500$.
(c) Differential current cross correlation at zero temperature for the representative above-EP point $|\delta\kappa|/t=1.00$. The dotted lines mark the inherited CAR--ECT balance points $V_c^{\rm low}$ and $V_c^{\rm high}$, while the dash-dotted line marks the additional noise-specific zero $V_n$. Throughout, $t=1$, $\Delta=0.9t$, and $\bar\kappa=\bar\kappa_c=\sqrt{2t\Delta}$.
}
\label{fig:Fig3}
\end{figure}

{\it Observable critical bifurcation}---The EP-anchored interference bifurcation survives in directly measurable transport observables at three complementary levels: its sign topology in nonlocal conductance, its thermal observability scale, and its fluctuation structure in current correlations.   
We adopt the one-sided bias convention $V_L=0$ and $V_R=V$, with the superconductor grounded. 
Within the present two-channel setting, $G_{\rm NL}>0$ corresponds to CAR-dominated transport and $G_{\rm NL}<0$ to ECT-dominated transport. At particle-hole symmetry,
\begin{equation}
G_{\rm NL}(V,T)
=
\frac{e^2}{h}\int d\omega\,
[T_{\rm CAR}-T_{\rm ECT}]
[-\partial_\omega f(\omega-eV)] ,
\label{eq:GNL}
\end{equation}
where $f(\epsilon)=[e^{\epsilon/(k_BT)}+1]^{-1}$ is the Fermi function. At zero temperature, \(G_{\rm NL}(V,0)/e^2/h=T_{\rm CAR}(eV)-T_{\rm ECT}(eV)\), so the interference bifurcation of Fig.~\ref{fig:Fig2} is mapped directly onto the sign structure of the nonlocal conductance.

Figure~\ref{fig:Fig3}(a) shows this correspondence across the synchronized $H_-$ exceptional point. Below the EP, $G_{\rm NL}$ remains negative for all biases. At the EP it touches zero tangentially at the coalesced interference boundary, while above the EP it becomes positive over a finite bias interval. 
The bifurcation therefore appears directly as \(G_{\rm NL}<0 \rightarrow G_{\rm NL}=0 \rightarrow G_{\rm NL}>0\), corresponding to an ECT-only spectrum, critical touching, and an ECT–CAR–ECT sequence. 

The finite-temperature response retains the same critical origin. We define $T^*$ as the temperature at which the positive-conductance interval disappears. Close to the EP, thermal convolution of the locally parabolic interference profile gives~\cite{sm}
\begin{equation}
k_BT^*
\simeq
\sqrt{
\frac{|J_-|}{2\pi^2}
\left(
|\delta\kappa|-|\delta\kappa|_{\rm EP}
\right)}
\propto
\left(
|\delta\kappa|-|\delta\kappa|_{\rm EP}
\right)^{1/2}.
\label{eq:Tstar-scaling}
\end{equation}
Thus the thermal observability scale inherits the same critical exponent, \(\beta_T = \beta = 1/2\). The numerical results in Fig.~\ref{fig:Fig3}(b) follow this square-root onset near the bifurcation. 
An independent free-exponent fit to the near-EP numerical data, $k_BT^*/t=A_{T,{\rm fit}}[(|\delta\kappa|-|\delta\kappa|_{\rm EP})/t]^{\beta_{T,{\rm fit}}}$, gives $\beta_{T,{\rm fit}}=0.500$, in agreement with the analytical prediction  \(\beta_T = \beta = 1/2\). 
Therefore the same square-root criticality governs both the zero-temperature opening of the CAR window and the thermal scale over which it remains observable. 

Current fluctuations provide a complementary probe of the same interference structure~\cite{MortenBelzig08prb}. At zero temperature under the same one-sided bias protocol, \(dS_{LR}(eV)/d(eV) = (e^2/h) \mathcal S_{LR}(eV)\), where $\mathcal S_{LR}(\omega)$ is the energy-resolved cross-noise density. For the lead-induced realization, it factorizes exactly as
\begin{equation}
\mathcal S_{LR}(\omega)
=
\mathcal F_{\rm int}(\omega)
\mathcal R_{\rm noise}(\omega),
\label{eq:noise-factorization}
\end{equation}
where $\mathcal F_{\rm int}$ is the interference factor that determines the CAR--ECT balance (Supplemental Material~\cite{sm}). 
Consequently, the two CAR--ECT balance boundaries $V_c^{\rm low}$ and $V_c^{\rm high}$ are necessarily inherited as zeros of the differential cross correlation, whereas additional zeros of $\mathcal R_{\rm noise}$, such as \(V_n\) in Fig.~\ref{fig:Fig3}(c), occur without a CAR--ECT balance and therefore reflect fluctuation-specific structure not present in the nonlocal conductance. The conductance and noise therefore expose complementary manifestations of the same EP-anchored interference bifurcation.

These observable signatures therefore probe the synchronized intersection selected in Fig.~\ref{fig:Fig2}(a); away from synchronization, the same interference bifurcation persists but is displaced from the spectral EP according to the control-space unfolding shown there and detailed in the Supplemental Material~\cite{sm}.

As an independent spectral cross-check, the sector-selective exceptional point underlying the synchronized bifurcation exhibits the Jordan-critical conditional dynamics derived in the Supplemental Material~\cite{sm}.

{\it Experimental feasibility}---The coherent and transport ingredients underlying the predicted bifurcation are available in hybrid semiconductor--superconductor quantum-dot devices, where tunable ECT and CAR couplings together with local and nonlocal transport spectroscopy have been demonstrated~\cite{BordinKouwenhovenDvir23prx,BordinDvir24prl}.
In the minimal lead-induced realization, synchronization requires $\bar{\Gamma}^{N}=\sqrt{2t\Delta}$ and
$|\Gamma_L^{N}-\Gamma_R^{N}|=4|t-\Delta|$.
Alternatively, for greater independent control, the transport couplings may be kept fixed while engineered local loss controls the total linewidths. 
For symmetric transport couplings this gives $\Gamma^{N}+\bar{\gamma}=\sqrt{2t\Delta}$ and $|\delta\gamma|=4|t-\Delta|$. 
Experimentally, controlled synchronization can be verified by tuning the mean linewidth through \(\bar{\kappa}_c\) and observing the reversal of the relative ordering of the spectral and interference thresholds. At the synchronized point, the coalescence of the relevant spectral resonances coincides with that of the CAR--ECT boundaries, followed beyond the EP by the opening of the positive nonlocal-conductance window; implementation details and detuning robustness are given in the Supplemental Material~\cite{sm}.

{\it Concluding remarks}---We have shown that a spectral exceptional point and an independently defined interference criticality can be deliberately synchronized at a controllable intersection. In the superconducting realization studied here, a sector-selective EP reorganizes interference between Bogoliubov propagation sectors so that an ECT-only spectrum develops a finite CAR-dominated window. 
Away from synchronization, tuning the mean linewidth reverses the relative ordering of the two distinct critical structures.  
At the synchronized point, the CAR window is born at the EP and opens with a square-root critical law. The intersection can moreover be restored under weak sector-mixing detuning, demonstrating that the synchronization is controlled rather than generic. The resulting bifurcation is directly visible in nonlocal conductance, the same critical exponent governs the temperature scale over which the CAR regime remains observable, and current correlations provide a complementary fluctuation fingerprint.

More broadly, these results extend the role of exceptional points beyond spectral singularities and sources of anomalous response. They show that distinct spectral and interference critical structures can be organized in a common control space and deliberately brought into coincidence. An EP can thereby serve as a spectral control point for an independently defined interference criticality, converting spectral coalescence into a qualitative bifurcation of coherent pathways. 
Superconducting transport provides a directly observable realization of this spectral–interference mechanism.  
This perspective suggests non-Hermitian spectral engineering as a route to organizing, synchronizing, and controlling critical interference phenomena in quantum systems.

{\it Acknowledgments}---This work is supported by National Key Research and Development Program of China (2025YFE0217400).

{\it Data availability}---The data is available from the corresponding authors upon reasonable request.

\end{document}


\title{Supplemental Material for ``Synchronizing Spectral and Interference Criticalities at an Exceptional Point''}

\author{Zeng-Zhao Li}
\email{lizengzhao@iqasz.cn}
\affiliation{International Quantum Academy, Shenzhen, 518048, China}

\author{Xiao Xue}
\email{xiao.xue@ustc.edu.cn}
\affiliation{International Quantum Academy, Shenzhen, 518048, China}
\affiliation{Hefei National Laboratory, University of Science and Technology of China, Hefei 230088, China}


\begin{abstract}
This Supplemental Material provides the derivation of the effective Bogoliubov-sector Hamiltonian and establishes the analytic separation, synchronization, and unfolding of the spectral and interference criticalities. It further presents robustness and experimental-implementation analyses, finite-temperature and current-correlation transport formulas, and the Jordan-critical conditional dynamics associated with the synchronized exceptional point.
\end{abstract}

\date{\today}
\maketitle
\tableofcontents

\section{Derivation of the retarded effective Hamiltonian}
\label{sec:SM-Hamiltonian}

\subsection{Integrating out the reservoirs}
\label{sec:SM-reservoirs}

We briefly derive the retarded effective Hamiltonian used in the main text.  The coherent superconducting subsystem is coupled locally to normal transport reservoirs and, optionally, to auxiliary loss channels.  The total Hamiltonian may be written as
\begin{equation}
H = H_{\rm sys} + \sum_{\alpha=L,R}
\left(
H_{\alpha}^{N}
+
H_{\alpha}^{\rm aux}
+
H_{\alpha}^{T}
\right),
\label{eq:SM-total-H}
\end{equation}
where \(H_{\rm sys}\) denotes the coherent two-site Hamiltonian representing the superconducting double-quantum-dot realization of the main text.
For a normal reservoir,
\begin{align}
H_{\alpha}^{N}
&=
\sum_k
\epsilon_{\alpha k}
c_{\alpha k}^{\dagger}c_{\alpha k},
\\
H_{\alpha}^{T}
&=
\sum_k
\left(
v_{\alpha k}
c_{\alpha k}^{\dagger}d_\alpha
+\mathrm{H.c.}
\right),
\label{eq:SM-lead-coupling}
\end{align}
with an analogous local coupling to the auxiliary reservoir.

Integrating out the reservoirs gives the retarded Green function of the central superconducting subsystem,
\begin{equation}
G^R(\omega)
=
\left[
\omega+i0^+
-
H_{\rm BdG}
-
\Sigma_N^R(\omega)
-
\Sigma_{\rm aux}^R(\omega)
\right]^{-1}.
\label{eq:SM-Dyson}
\end{equation}
In the wide-band limit, the real parts of the reservoir self-energies can be absorbed into the local energies, while their imaginary parts become energy independent.  For local particle-hole-symmetric couplings,
\begin{align}
\Sigma_N^R
&=
-\frac{i}{2}
\sum_{\alpha=L,R}
\Gamma_\alpha^N P_\alpha,
\\
\Sigma_{\rm aux}^R
&=
-\frac{i}{2}
\sum_{\alpha=L,R}
\gamma_\alpha P_\alpha ,
\label{eq:SM-selfenergies}
\end{align}
where \(P_\alpha\) projects onto the local Nambu subspace of site \(\alpha\).  
Defining the total linewidth
\begin{equation}
\kappa_\alpha
=
\Gamma_\alpha^N+\gamma_\alpha ,
\label{eq:SM-total-linewidth}
\end{equation}
Eq.~(\ref{eq:SM-Dyson}) becomes
\begin{equation}
G^R(\omega)
=
\left[
\omega-H_{\rm eff}^{R}
\right]^{-1},
\qquad
H_{\rm eff}^{R}
=
H_{\rm BdG}
-\frac{i}{2}
\sum_{\alpha=L,R}
\kappa_\alpha P_\alpha .
\label{eq:SM-Heff-derived}
\end{equation}

This form makes explicit the distinction emphasized in the main text. Both the normal leads and the auxiliary reservoirs contribute to the spectral linewidth \(\kappa_\alpha\), whereas only the normal-lead couplings \(\Gamma_\alpha^N\) enter the transport injection and collection factors.  The exceptional-point condition can therefore be controlled either through the transport couplings themselves or, independently, through auxiliary loss.

\subsection{Canonical BdG representation and transformation to the Bogoliubov-sector basis}
\label{sec:SM-canonical-BdG}

We derive here the relation between the canonical fermionic Bogoliubov--de Gennes representation and the working basis used in the main text.  We consider the resonant minimal two-site model
\begin{equation}
H_{\rm sys}
=
t\left(
d_L^\dagger d_R+d_R^\dagger d_L
\right)
-\Delta
\left(
d_L^\dagger d_R^\dagger+d_R d_L
\right),
\label{eq:SM-Hsys}
\end{equation}
where \(t\) is the normal intersite hopping and \(\Delta\) is the nonlocal pairing amplitude.  The overall sign and phase convention of \(\Delta\) can be changed by a gauge transformation and does not affect the physical results.

In the canonical Nambu basis \(\Phi_c = \left(d_L,\,d_R,\,d_L^\dagger,\,d_R^\dagger\right)^T \), 
Eq.~(\ref{eq:SM-Hsys}) can be written as
\begin{equation}
H_{\rm sys}
=
\frac{1}{2}
\Phi_c^\dagger
\mathcal H_c
\Phi_c
+\mathrm{const.},
\label{eq:SM-BdG-quadratic}
\end{equation}
with
\begin{equation}
\mathcal H_c
=
\begin{pmatrix}
0&t&0&-\Delta\\
t&0&\Delta&0\\
0&\Delta&0&-t\\
-\Delta&0&-t&0
\end{pmatrix}.
\label{eq:SM-Hcanonical}
\end{equation}
The anomalous block is antisymmetric, as required for spinless fermionic pairing, 
\begin{equation}
\Delta_c
=
\begin{pmatrix}
0&-\Delta\\
\Delta&0
\end{pmatrix},
\qquad
\Delta_c^T=-\Delta_c .
\label{eq:SM-pairing-antisymmetry}
\end{equation} 
The canonical BdG Hamiltonian obeys particle-hole symmetry.  Introducing the antiunitary operator \(\mathcal C_c = \rho_x K\), where \(\rho_x\) exchanges particle and hole blocks and \(K\) denotes complex conjugation, one finds \(\mathcal C_c \mathcal H_c \mathcal C_c^{-1} = -\mathcal H_c\) with \(\mathcal C_c^2=+1\). 

For the sector decomposition used in the main text it is more convenient to first reorder the basis locally by site. We define \(\Phi_{\rm loc} = (d_L,\,d_L^\dagger,\,d_R,\,d_R^\dagger)^T\) related to the canonical Nambu basis by
\begin{equation}
\Phi_c
=
P\Phi_{\rm loc},
\qquad
P=
\begin{pmatrix}
1&0&0&0\\
0&0&1&0\\
0&1&0&0\\
0&0&0&1
\end{pmatrix}.
\label{eq:SM-permutation}
\end{equation}
The corresponding Hamiltonian is
\begin{equation}
\mathcal H_{\rm loc}
=
P^\dagger
\mathcal H_c
P
=
\begin{pmatrix}
0&0&t&-\Delta\\
0&0&\Delta&-t\\
t&\Delta&0&0\\
-\Delta&-t&0&0
\end{pmatrix}.
\label{eq:SM-Hlocal}
\end{equation}

We next perform a local Nambu phase transformation on the right-hole basis state, \(\)
\begin{equation}
|R,h\rangle
\rightarrow
-|R,h\rangle ,
\label{eq:SM-hole-phase}
\end{equation}
implemented by
\begin{equation}
U
=
\mathrm{diag}
\left(
1,1,1,-1
\right).
\label{eq:SM-Uphase}
\end{equation}
The working Nambu spinor is therefore
\begin{equation}
\Psi
=
U^\dagger\Phi_{\rm loc}
=
\left(
d_L,\,
d_L^\dagger,\,
d_R,\,
-d_R^\dagger
\right)^T .
\label{eq:SM-working-spinor}
\end{equation}
The coherent Hamiltonian in this basis becomes
\begin{align}
\mathcal H_{\rm w}
&=
U^\dagger
\mathcal H_{\rm loc}
U
=
\begin{pmatrix}
0&0&t&\Delta\\
0&0&\Delta&t\\
t&\Delta&0&0\\
\Delta&t&0&0
\end{pmatrix}.
\label{eq:SM-Hworking-matrix}
\end{align}
Introducing Pauli matrices \(\tau_i\) in the site space \(\{L,R\}\) and \(\sigma_i\) in the local Nambu space \(\{e,h\}\), Eq.~(\ref{eq:SM-Hworking-matrix}) takes the compact form
\begin{equation}
\mathcal H_{\rm w}
=
\tau_x\otimes
\left(
t\sigma_0+\Delta\sigma_x
\right).
\label{eq:SM-Hworking}
\end{equation}
This is the working representation employed in the main text.

The particle-hole operator transforms together with the basis. In the local site--Nambu ordering, its unitary part is \(\mathcal C_{\rm loc} = \left( \tau_0\otimes\sigma_x \right)K\). 
After the phase transformation of Eq.~(\ref{eq:SM-Uphase}), the particle-hole operator becomes
\begin{align}
\mathcal C_{\rm w}
&=
U^\dagger
\left(
\tau_0\otimes\sigma_x
\right)
U^* K
=
\left(
\tau_z\otimes\sigma_x
\right)K .
\label{eq:SM-Cworking}
\end{align}
One directly verifies \(\mathcal C_{\rm w} \mathcal H_{\rm w} \mathcal C_{\rm w}^{-1} = -\mathcal H_{\rm w}\) with \(\mathcal C_{\rm w}^2=+1\). 
Thus the symmetric-looking anomalous coupling in Eq.~(\ref{eq:SM-Hworking}) is entirely a consequence of the Nambu-basis choice; the fermionic BdG particle-hole structure is unchanged.

The same particle-hole constraint is preserved after including local broadening. With
\begin{equation}
H_{\rm eff}^{R}
=
\mathcal H_{\rm w}
-\frac{i}{2}
\sum_{\alpha=L,R}
\kappa_\alpha P_\alpha ,
\label{eq:SM-Heff-general}
\end{equation}
where \(P_L = \frac{\tau_0+\tau_z}{2}\otimes\sigma_0\) and \(P_R = \frac{\tau_0-\tau_z}{2} \otimes\sigma_0\),
the loss term is proportional to the identity in the local Nambu subspace. Since complex conjugation changes \(-i\kappa_\alpha/2\) into \(+i\kappa_\alpha/2\), one obtains \(\mathcal C_{\rm w} H_{\rm eff}^{R} \mathcal C_{\rm w}^{-1} = -H_{\rm eff}^{R}\). 
Accordingly, the complex spectrum satisfies the BdG pairing \(E \longleftrightarrow -E^*\). 

Finally, the propagation-sector decomposition follows directly from the Nambu structure of Eq.~(\ref{eq:SM-Hworking}).  The eigenstates of \(\sigma_x\) are
\begin{equation}
|\pm_{\rm N}\rangle
=
\frac{|e\rangle\pm|h\rangle}{\sqrt{2}},
\qquad
\sigma_x|\pm_{\rm N}\rangle
=
\pm|\pm_{\rm N}\rangle .
\label{eq:SM-Nambu-eigenstates}
\end{equation}
Transforming to the basis \(\{|L,+_{\rm N}\rangle, |R,+_{\rm N}\rangle, |L,-_{\rm N}\rangle, |R,-_{\rm N}\rangle\}\) therefore block diagonalizes the effective Hamiltonian,
\begin{equation}
H_{\rm eff}^{R}
\longrightarrow
H_+\oplus H_- ,
\label{eq:SM-block-decomposition}
\end{equation}
with
\begin{equation}
H_\pm
=
-\frac{i\bar\kappa}{2}\mathbb I
+
J_\pm\tau_x
-i\frac{\delta\kappa}{4}\tau_z ,
\qquad
J_\pm=t\pm\Delta ,
\label{eq:SM-Hpm}
\end{equation}
where \(\bar\kappa = \frac{\kappa_L+\kappa_R}{2}\) and \(\delta\kappa = \kappa_L-\kappa_R\). 
The two complex eigenvalues within sector \(s=\pm\) are
\begin{equation}
E_{s,\eta}
=
-\frac{i\bar\kappa}{2}
+
\eta
\sqrt{
J_s^2-\frac{\delta\kappa^2}{16}
},
\qquad
\eta=\pm1 ,
\label{eq:SM-sector-spectrum}
\end{equation}
and therefore coalesce at
\begin{equation}
|\delta\kappa|_{\rm EP}^{(s)}
=
4|J_s|.
\label{eq:SM-sector-EP}
\end{equation}
Hence the effective couplings \(J_+=t+\Delta\) and \(J_-=t-\Delta\) are not introduced phenomenologically: they are the exact propagation eigen-couplings of the two coherent local Nambu combinations.

\section{Spectral and Bogoliubov-interference criticalities}
\label{sec:SM-interference}

We derive here the analytic structure underlying the CAR--ECT interference discussed in the main text and clarify its relation to the sector-selective exceptional point. The central distinction is that the exceptional point is a spectral singularity of an individual Bogoliubov propagation sector, whereas the CAR--ECT balance defines an interference criticality through the relative complex response of both sectors. The associated spectral and interference critical structures are therefore generically distinct, but their relative position can be tuned continuously and brought into coincidence at the synchronized operating point studied in Figs.~2 and 3 of the main text.

\subsection{Analytic CAR--ECT interference boundaries}
\label{sec:SM-interference-boundaries}

For each Bogoliubov sector $s=\pm$, the nonlocal propagator introduced in the main text is
\begin{equation}
G_s(\omega)
=
\frac{J_s}{D_s(\omega)},
\qquad
D_s(\omega)
=
(\omega+i\kappa_L/2)(\omega+i\kappa_R/2)-J_s^2 .
\label{eq:SM-Gs}
\end{equation}
Using \(\bar\kappa = \frac{\kappa_L+\kappa_R}{2}\) and \(\delta\kappa=\kappa_L-\kappa_R\), 
we write
\begin{equation}
D_s(\omega)
=
\omega^2-\Lambda_s+i\bar\kappa\omega ,
\qquad
\Lambda_s = J_s^2+\frac{\kappa_L\kappa_R}{4}
= J_s^2+\frac{\bar\kappa^2}{4} -\frac{\delta\kappa^2}{16}.
\label{eq:SM-Ds}
\end{equation}

Transforming the sector propagators back to the physical electron--hole channels gives
\begin{equation}
G_{ee}^{LR}
=
\frac{G_++G_-}{2},
\qquad
G_{eh}^{LR}
=
\frac{G_+-G_-}{2}.
\label{eq:SM-physical-amplitudes}
\end{equation}
Here the right-hole phase convention introduced in Eq.~(\ref{eq:SM-hole-phase}) does not alter these relations for the transport process considered here, since the incoming state is the physical right-electron state. The corresponding nonlocal probabilities are therefore
\begin{equation}
T_{\rm ECT}
=
\Gamma_L^N\Gamma_R^N |G_{ee}^{LR}|^2,
\qquad
T_{\rm CAR}
=
\Gamma_L^N\Gamma_R^N |G_{eh}^{LR}|^2,
\label{eq:SM-physical-transmissions}
\end{equation}
and hence
\begin{equation}
T_{\rm CAR}-T_{\rm ECT}
=
-\Gamma_L^N\Gamma_R^N\,
\operatorname{Re}(G_+G_-^*).
\label{eq:SM-interference-identity}
\end{equation}
We use the normalized CAR--ECT contrast \(\mathcal{C}_{\rm CE}=(T_{\rm CAR}-T_{\rm ECT})/(T_{\rm CAR}+T_{\rm ECT})\), for which \(\mathcal{C}_{\rm CE}=0\) marks the CAR--ECT balance boundary.  
The interference quantity controlling the difference between CAR and ECT is
\begin{align}
\operatorname{Re}
\!\left[
G_+(\omega)G_-^*(\omega)
\right]
&=
J_+J_-
\operatorname{Re}
\left[
\frac{1}{D_+(\omega)D_-^*(\omega)}
\right]
=
J_+J_-
\frac{{\cal F}_{\rm int}(\omega)}
{|D_+(\omega)D_-(\omega)|^2},
\label{eq:SM-interference-factor}
\end{align}
where
\begin{equation}
{\cal F}_{\rm int}(\omega)
=
(\omega^2-\Lambda_+)(\omega^2-\Lambda_-)
+
\bar\kappa^2\omega^2 .
\label{eq:SM-Fomega}
\end{equation}

For the parameter regime considered in the main text,
$t>\Delta>0$, one has
\begin{equation}
J_+=t+\Delta>0,
\qquad
J_-=t-\Delta>0 ,
\label{eq:SM-Jsign}
\end{equation}
and therefore $J_+J_->0$.  Since the denominator in Eq.~(\ref{eq:SM-interference-factor}) is positive away from poles, the sign of the interference is determined entirely by ${\cal F}_{\rm int}(\omega)$. Using \(T_{\rm CAR}-T_{\rm ECT} = -\Gamma_L^N\Gamma_R^N \operatorname{Re} \!\left(G_+G_-^*\right)\) in Eq.~(\ref{eq:SM-interference-identity}), 
the CAR--ECT balance condition is consequently
\begin{equation}
{\cal F}_{\rm int}(\omega)=0 .
\label{eq:SM-balance-condition}
\end{equation}

Introducing \(x=\omega^2\), Eq.~(\ref{eq:SM-balance-condition}) becomes
\begin{equation}
x^2
-
\left(
\Lambda_+ + \Lambda_- - \bar\kappa^2
\right)x
+
\Lambda_+\Lambda_-
=
0 .
\label{eq:SM-interference-quadratic}
\end{equation}
The two roots are
\begin{equation}
x_\pm
=
\frac{
\Lambda_+ + \Lambda_- - \bar\kappa^2
\pm\sqrt{\mathcal D_{\rm int}}
}{2},
\label{eq:SM-xroots}
\end{equation}
where
\begin{equation}
\mathcal D_{\rm int}
=
\left(
\Lambda_+ + \Lambda_- - \bar\kappa^2
\right)^2
-
4\Lambda_+\Lambda_- .
\label{eq:SM-Dint-basic}
\end{equation}
Whenever both roots are real and positive, the positive-energy interference boundaries are
\begin{equation}
\omega_c^{\rm low}
=
\sqrt{x_-},
\qquad
\omega_c^{\rm high}
=
\sqrt{x_+}.
\label{eq:SM-omegac}
\end{equation}
Particle-hole symmetry gives the corresponding boundaries at negative energies.

The existence of two distinct positive boundaries requires
\begin{equation}
\mathcal D_{\rm int}>0,
\qquad
\Lambda_+ + \Lambda_- - \bar\kappa^2>0,
\qquad
\Lambda_+\Lambda_->0 .
\label{eq:SM-boundary-conditions}
\end{equation}
For $\Lambda_\pm>0$, the last condition is automatic. The discriminant can
also be factorized as
\begin{equation}
\mathcal D_{\rm int}
=
\left[
\left(\sqrt{\Lambda_+}-\sqrt{\Lambda_-}\right)^2
-\bar\kappa^2
\right]
\left[
\left(\sqrt{\Lambda_+}+\sqrt{\Lambda_-}\right)^2
-\bar\kappa^2
\right].
\label{eq:SM-D-factorized}
\end{equation}
This form makes explicit that the appearance of two separate interference crossings is controlled by the competition between the separation of the two Bogoliubov-sector scales and their common broadening.

Because ${\cal F}_{\rm int}(\omega)$ is an upward-opening quadratic in $x=\omega^2$, for two positive roots $x_-<x_+$ one has
\begin{equation}
{\cal F}_{\rm int}(\omega)
\begin{cases}
>0, & \omega^2<x_- ,\\
<0, & x_-<\omega^2<x_+ ,\\
>0, & \omega^2>x_+ .
\end{cases}
\label{eq:SM-F-sign}
\end{equation}
For $J_+J_->0$, Eq.~(\ref{eq:SM-interference-identity}) therefore gives \(\mathrm{ECT}\rightarrow\mathrm{CAR}\rightarrow\mathrm{ECT}\)
as energy is swept through the two positive interference boundaries.
The finite CAR-dominated window thus follows analytically from the two-sector interference structure rather than from a particular numerical parameter choice.

It is important to distinguish these interference boundaries from the exceptional-point condition. The latter is a spectral property of an individual Bogoliubov sector, $|\delta\kappa|_{\rm EP}^{(s)}=4|J_s|$ [Eq.~(\ref{eq:SM-sector-EP})], whereas Eqs.~(\ref{eq:SM-balance-condition}) and (\ref{eq:SM-omegac}) depend on the relative complex response of both sectors.  The spectral EP can therefore reorganize the interference landscape by restructuring one sector propagator, but is not itself, in general, a CAR--ECT balance boundary.

\subsection{Generic separation of the spectral and interference criticalities}
\label{sec:SM-generic-separation}

\begin{figure}[t]
\centering
\includegraphics[width=0.46\columnwidth]{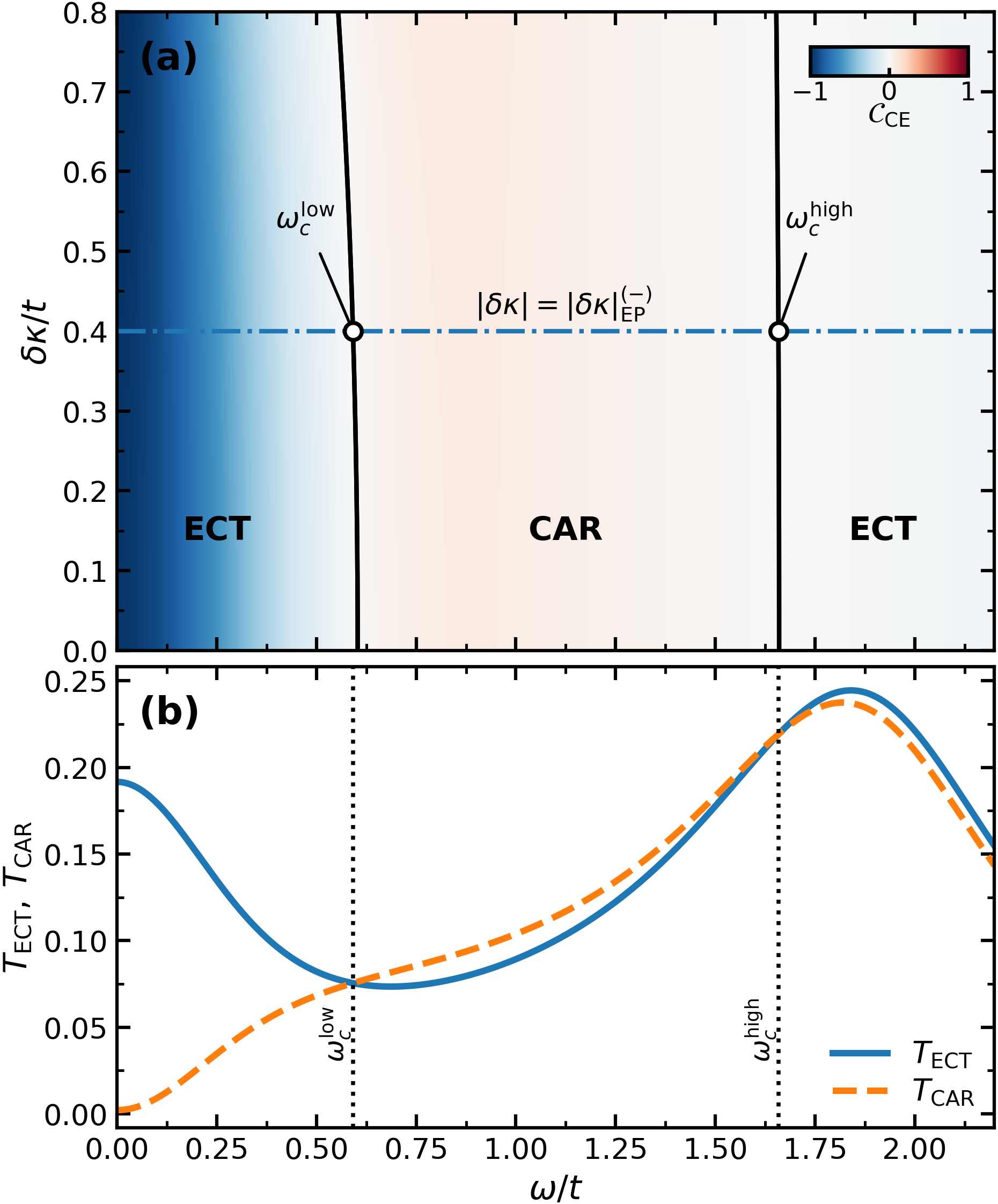}
\caption{
\textbf{Generic separation of the spectral and interference
criticalities.}
(a) CAR--ECT contrast $C_{\rm CE}$ versus energy $\omega$ and loss
imbalance $|\delta\kappa|$ for the unsynchronized choice
$\bar\kappa=t$.
The black solid curves show the analytic CAR--ECT balance boundaries,
$\mathcal{C}_{\rm CE}=0$, while the blue dash-dotted horizontal line marks the
$H_-$ exceptional-point condition
$|\delta\kappa|_{\rm EP}^{(-)}=4|J_-|$.
At the spectral EP the two interference boundaries remain separated,
demonstrating that the EP and the interference bifurcation are
generically distinct.
(b) Cut at
$|\delta\kappa|=|\delta\kappa|_{\rm EP}^{(-)}$,
showing the resulting ECT--CAR--ECT sequence.
Parameters are $t=1$, $\Delta=0.9t$, and $\bar\kappa=t$.
}
\label{fig:SM-generic}
\end{figure}

The distinction above is illustrated explicitly in Fig.~\ref{fig:SM-generic}. Here the mean linewidth is chosen as $\bar\kappa=t$, away from the synchronized value \(\bar\kappa_c=\sqrt{2t\Delta}\). 
For the parameters $t=1$ and $\Delta=0.9t$, the $H_-$ sector reaches its exceptional point at
\begin{equation}
|\delta\kappa|_{\rm EP}^{(-)}
=
4|J_-|
=
0.4t .
\label{eq:SM-generic-EP}
\end{equation}
At this loss imbalance, however, the two CAR--ECT interference boundaries are already separated, so the spectral EP does not mark the birth of the CAR-dominated window.

Thus, in the generic unsynchronized situation, crossing the spectral EP changes the complex response of the $H_-$ sector and thereby reshapes the interference landscape, but it need not create or annihilate the CAR window itself. 
This provides a useful reference for the synchronized case studied in the main text, where the interference-critical manifold is tuned to intersect the spectral-EP condition, so that the two interference boundaries coalesce precisely at the spectral EP.

\subsection{Synchronization and two-parameter unfolding}
\label{sec:SM-critical-unfolding}

We now determine the condition under which the spectral EP and the CAR--ECT interference bifurcation coincide, and how this coincidence unfolds when the mean linewidth is detuned.

We write
\begin{equation}
d\equiv|\delta\kappa|, 
\qquad 
\Delta_J \equiv J_+^2-J_-^2>0 ,
\label{eq:SM-unfold-def}
\end{equation}
and recall that the exceptional point of the $H_-$ sector occurs at \(d=d_{\rm EP}\) with \(d_{\rm EP}=4|J_-|\). 
Importantly, the position of this spectral EP is independent of the mean linewidth $\bar\kappa$.

Using Eq.~(\ref{eq:SM-Dint-basic}), the interference discriminant can be written explicitly as
\begin{equation}
\mathcal D_{\rm int}
=
\left(J_+^2-J_-^2\right)^2
-
2\bar\kappa^2
\left(J_+^2+J_-^2\right)
+
\frac{\bar\kappa^2d^2}{4}.
\label{eq:SM-unfold-Dint}
\end{equation}
The interference bifurcation occurs when $\mathcal D_{\rm int}=0$.

Requiring the interference bifurcation to occur precisely at the $H_-$ exceptional point, $d=d_{\rm EP}$, we have \({\cal D}_{\rm int}|_{d=d_{\rm EP}}=\Delta_J(\Delta_J-2\bar{\kappa}^2)\). Since $\Delta_J>0$, the synchronization condition \({\cal D}_{\rm int}=0\) gives
\begin{equation}
\bar\kappa_c^2
=
\frac{\Delta_J}{2}
=
\frac{J_+^2-J_-^2}{2}
=
2t\Delta .
\label{eq:SM-unfold-kappac}
\end{equation}
At synchronization, the two interference roots coalesce. Using Eq.~(\ref{eq:SM-xroots}) gives 
\begin{equation}
x_+=x_-\equiv x_*
=\left.\frac{\Lambda_+ + \Lambda_- -\bar{\kappa}^2}{2}
\right|_{\bar{\kappa}=\bar{\kappa}_c,\; d=d_{\rm EP}}
=
\frac{3}{8}\Delta_J .
\label{eq:SM-bifurcation-x}
\end{equation}
Hence the positive-energy bifurcation point is 
\begin{equation}
\omega_*=\sqrt{x_*}=\sqrt{\frac{3\Delta_J}{8}}=\sqrt{\frac{3t\Delta}{2}} .
\label{eq:SM-bifurcation-omega}
\end{equation}
Thus the synchronized bifurcation occurs at \((\bar{\kappa},d,\omega)=(\bar{\kappa}_c,d_{\rm EP},\omega_*)\), together with the particle-hole-related point at \(-\omega_*\).

Equation~(\ref{eq:SM-unfold-Dint}) can then be rearranged exactly as
\begin{equation}
\mathcal D_{\rm int}
=
\frac{\bar\kappa^2}{4}
\left(
d^2-d_{\rm EP}^2
\right)
-
2\Delta_J
\left(
\bar\kappa^2-\bar\kappa_c^2
\right).
\label{eq:SM-unfold-Dint-rearranged}
\end{equation}
This form explicitly separates the displacement from the spectral EP, $d^2-d_{\rm EP}^2$, from the detuning of the mean linewidth, $\bar\kappa^2-\bar\kappa_c^2$.

Setting $\mathcal D_{\rm int}=0$ gives the interference-critical curve
\begin{equation}
d_{\rm int}^2(\bar\kappa)
=
d_{\rm EP}^2
+
\frac{8\Delta_J}{\bar\kappa^2}
\left(
\bar\kappa^2-\bar\kappa_c^2
\right),
\label{eq:SM-unfold-dint}
\end{equation}
or equivalently,
\begin{equation}
d_{\rm int}^2(\bar\kappa)
=
8\left(J_+^2+J_-^2\right)
-
\frac{
4\left(J_+^2-J_-^2\right)^2
}{
\bar\kappa^2
}.
\label{eq:SM-unfold-dint-alt}
\end{equation}
The spectral and interference criticalities therefore define distinct curves in the $(\bar\kappa,d)$ control plane.  They intersect at \((\bar\kappa,d) = (\bar\kappa_c,d_{\rm EP})\). 

The interference-critical curve enters the physical $d\geq0$ plane at
\begin{equation}
\bar\kappa_0^2
=
\frac{
\Delta_J^2
}{
2\left(J_+^2+J_-^2\right)
},
\qquad
d_{\rm int}(\bar\kappa_0)=0 .
\label{eq:SM-unfold-kappa0}
\end{equation}
The resulting ordering of the spectral and interference criticalities is
\begin{equation}
\begin{array}{lll}
\bar\kappa<\bar\kappa_0
&\Longrightarrow&
\mathcal D_{\rm int}>0
\quad\text{already at }d=0,
\\[2pt]
\bar\kappa_0<\bar\kappa<\bar\kappa_c
&\Longrightarrow&
0<d_{\rm int}<d_{\rm EP},
\\[2pt]
\bar\kappa=\bar\kappa_c
&\Longrightarrow&
d_{\rm int}=d_{\rm EP},
\\[2pt]
\bar\kappa>\bar\kappa_c
&\Longrightarrow&
d_{\rm int}>d_{\rm EP}.
\end{array}
\label{eq:SM-unfold-ordering}
\end{equation}
For $\bar\kappa<\bar\kappa_0$, two real interference boundaries are already present at zero loss imbalance and remain present as $d$ is increased. 
For $\bar\kappa_0<\bar\kappa<\bar\kappa_c$, the interference bifurcation occurs at a finite imbalance before the $H_-$ sector reaches its spectral EP. 
For $\bar\kappa>\bar\kappa_c$, the spectral EP is crossed first, whereas the interference bifurcation occurs only at a larger imbalance. 
At $\bar\kappa=\bar\kappa_c$, the two critical conditions coincide. Thus the synchronized point is their deliberately tunable intersection rather than a generic identification of spectral and interference criticality.

The local structure near the synchronized intersection can be made explicit. Let \(d=d_{\rm EP}+\delta d\) and \(\bar\kappa^2 = \bar\kappa_c^2+\delta K\) with
$|\delta d|\ll d_{\rm EP}$ and $|\delta K|\ll\bar\kappa_c^2$.
Expanding Eq.~(\ref{eq:SM-unfold-Dint-rearranged}) to leading order gives
\begin{equation}
\mathcal D_{\rm int}
=
\Delta_J
\left(
|J_-|\,\delta d-2\delta K
\right)
+
O\!\left(
\delta d^2,\,
\delta d\,\delta K
\right).
\label{eq:SM-unfold-local-Dint}
\end{equation}
It is therefore useful to introduce the local unfolding coordinate \(\eta \equiv |J_-| \left(d-d_{\rm EP}\right) - 2\left(\bar\kappa^2-\bar\kappa_c^2\right)\). 
With this definition,
\begin{equation}
\mathcal D_{\rm int}
=
\Delta_J\eta
+
O\!\left(
\delta d^2,\,
\delta d\,\delta K
\right).
\label{eq:SM-unfold-eta-D}
\end{equation}
Thus, along a generic trajectory crossing the interference-critical manifold transversely near the synchronized point, the leading-order local interference structure is distinguished by
\begin{equation}
\eta<0:
\ \text{no real boundaries},
\qquad
\eta=0:
\ \text{double boundary},
\qquad
\eta>0:
\ \text{two real boundaries}.
\label{eq:SM-unfold-regimes}
\end{equation}

The same local expansion generalizes the square-root opening derived in the main text.  Along a generic transverse crossing of the interference-critical manifold near the synchronized point, the separation of the two positive-energy interference boundaries obeys
\begin{equation}
\Delta\omega_{\rm CAR}
\simeq
\sqrt{\frac{2}{3}\eta}
=
\sqrt{
\frac{2}{3}
\left[
|J_-|
\left(d-d_{\rm EP}\right)
-
2\left(
\bar\kappa^2-\bar\kappa_c^2
\right)
\right]
},
\label{eq:SM-unfold-window}
\end{equation}
for $\eta\rightarrow0^+$. On the synchronized cut
$\bar\kappa=\bar\kappa_c$, this reduces to
\begin{equation}
\Delta\omega_{\rm CAR}
\simeq
\sqrt{
\frac{2|J_-|}{3}
\left(
d-d_{\rm EP}
\right)
},
\label{eq:SM-unfold-window-sync}
\end{equation}
as quoted in the main text.

Equation~(\ref{eq:SM-unfold-window}) shows that the square-root opening is not restricted to the precisely synchronized trajectory. More generally, it is the local square-root opening associated with crossing the interference-critical manifold in the two-dimensional control space. Synchronization selects the operating point at which 
this manifold crosses the spectral EP.

For the parameters used in Figs.~2 and 3 of the main text,
$t=1$ and $\Delta=0.9t$, one has
\begin{equation}
J_+=1.9t,
\qquad
J_-=0.1t,
\qquad
d_{\rm EP}=0.4t,
\qquad
\bar\kappa_c=\sqrt{1.8}\,t ,
\label{eq:SM-unfold-numerical}
\end{equation}
together with
\begin{equation}
\bar\kappa_0
=
\frac{
\Delta_J
}{
\sqrt{2(J_+^2+J_-^2)}
}
\simeq
1.33793\,t .
\label{eq:SM-unfold-kappa0-numerical}
\end{equation}
As shown in Fig.~2(a) of the main text, the spectral- and interference-critical curves cross at the synchronized operating point.
Detuning $\bar\kappa$ continuously unfolds this intersection and reverses the ordering of the two criticalities across $\bar\kappa_c$.

\subsection{Lead-induced and auxiliary-loss realizations}
\label{sec:SM-loss-realizations}

The total linewidth entering the retarded sector propagators is $\kappa_\alpha=\Gamma_\alpha^N+\gamma_\alpha$ [Eq.~(\ref{eq:SM-total-linewidth})]. 
Consequently, the spectral quantities $D_s(\omega)$, $G_s(\omega)$, and the exceptional-point condition depend on the reservoirs only through $\kappa_L$ and $\kappa_R$.  The transport probabilities, however, retain the normal-lead prefactor $\Gamma_L^N\Gamma_R^N$.

In the minimal lead-induced realization used for the primary transport calculations, \(\gamma_L=\gamma_R=0\) and \(\kappa_\alpha=\Gamma_\alpha^N\). 
At fixed \(\bar\Gamma^N = (\Gamma_L^N+\Gamma_R^N)/2\), varying \(\delta\Gamma^N = \Gamma_L^N-\Gamma_R^N\) therefore tunes the non-Hermitian spectrum while simultaneously changing the transport prefactor.

Alternatively, the normal transport couplings may be held fixed while an asymmetric auxiliary loss produces the required linewidth imbalance. 
In this case, \(\delta\kappa = \delta\Gamma^N+\delta\gamma\) with \(\delta\gamma = \gamma_L-\gamma_R\), 
and for symmetric transport contacts
$\Gamma_L^N=\Gamma_R^N$,
\begin{equation}
\delta\kappa=\delta\gamma .
\label{eq:SM-aux-only}
\end{equation}
The normal-lead prefactor then remains unchanged as the system is tuned through the exceptional point.  Reproducing the CAR--ECT reorganization under this protocol therefore separates the non-Hermitian spectral control from a trivial change of contact transparency.

\subsection{Robustness to onsite-energy detuning}
\label{sec:SM-onsite-detuning}

The analytic sector decomposition used above is exact at the resonant operating point.  We now show that the exceptional-point-controlled interference is not restricted to this exactly solvable limit.

We add local onsite energies
\begin{equation}
H_\epsilon
=
\epsilon_L d_L^\dagger d_L
+
\epsilon_R d_R^\dagger d_R
\label{eq:SM-onsite-H}
\end{equation}
to the coherent Hamiltonian. 
In the working site--Nambu basis of Sec.~\ref{sec:SM-Hamiltonian}, the coherent Hamiltonian becomes
\begin{equation}
H_w(\epsilon_L,\epsilon_R)
=
\tau_x\otimes
(t\sigma_0+\Delta\sigma_x)
+
\left(
\bar\epsilon\,\tau_0
+
\frac{\delta\epsilon}{2}\tau_z
\right)
\otimes\sigma_z ,
\label{eq:SM-detuned-Hw}
\end{equation}
where \(\bar\epsilon = \frac{\epsilon_L+\epsilon_R}{2}\) and \(\delta\epsilon = \epsilon_L-\epsilon_R\). 
The corresponding retarded effective Hamiltonian is
\begin{equation}
H_{\rm eff}^R
=
H_w(\epsilon_L,\epsilon_R)
-
\frac{i\bar\kappa}{2}I
-
\frac{i\delta\kappa}{4}
\tau_z\otimes\sigma_0 .
\label{eq:SM-detuned-Heff}
\end{equation}

Because the onsite-energy term is proportional to $\sigma_z$, whereas the propagation sectors at resonance are eigenstates of $\sigma_x$, finite detuning couples the two Bogoliubov sectors.  
Thus the exact block decomposition $H_{\rm eff}^R\rightarrow H_+\oplus H_-$ used in the resonant analytic treatment no longer applies. The detuned problem therefore provides a direct test of whether the interference reorganization relies on exact sector conservation.

Two representative detuning directions can still be treated analytically at the level of the full $4\times4$ spectrum. 
For common detuning, \(\epsilon_L=\epsilon_R=\epsilon\),  
the four eigenvalues can be written as
\begin{equation}
E_{\eta,\xi}
=
-\frac{i\bar\kappa}{2}
+
\eta\sqrt{Q_\xi^{(c)}},
\qquad
\eta,\xi=\pm1 ,
\label{eq:SM-common-spectrum}
\end{equation}
with
\begin{equation}
Q_\xi^{(c)} = t^2+\Delta^2+\epsilon^2-a^2 + \xi\,2 \sqrt{ t^2\Delta^2+ \epsilon^2(t^2-a^2)} ,
\qquad
a=\frac{\delta\kappa}{4} .
\label{eq:SM-Q-common}
\end{equation}
The exceptional point continuously connected to the $H_-$ exceptional point at resonance is obtained from $Q_-^{(c)}=0$, giving
\begin{equation}
|\delta\kappa|_{\rm EP}^{(c)}
=
4
\left|
\sqrt{t^2-\epsilon^2}-\Delta
\right|,
\qquad
|\epsilon|<|t|.
\label{eq:SM-EP-common}
\end{equation}
Equation~(\ref{eq:SM-EP-common}) reduces to $|\delta\kappa|_{\rm EP}=4|t-\Delta|$ as $\epsilon\rightarrow0$.

For antisymmetric detuning, \(\epsilon_L=-\epsilon_R=\epsilon\), 
the spectrum takes the analogous form
\begin{equation}
E_{\eta,\xi}
=
-\frac{i\bar\kappa}{2}
+
\eta\sqrt{Q_\xi^{(a)}} ,
\label{eq:SM-antisym-spectrum}
\end{equation}
where
\begin{equation}
Q_\xi^{(a)}
=
t^2+\Delta^2+\epsilon^2-a^2
+
\xi\,2
\sqrt{
\Delta^2(t^2+\epsilon^2)
-a^2\epsilon^2
}.
\label{eq:SM-Q-antisym}
\end{equation}
The exceptional branch connected continuously to the resonant
$H_-$ exceptional point is therefore
\begin{equation}
|\delta\kappa|_{\rm EP}^{(a)}
=
4
\left|
t-\sqrt{\Delta^2-\epsilon^2}
\right|,
\qquad
|\epsilon|<|\Delta|.
\label{eq:SM-EP-antisym}
\end{equation}

\begin{figure}
\centering
\includegraphics[width=0.46\columnwidth]{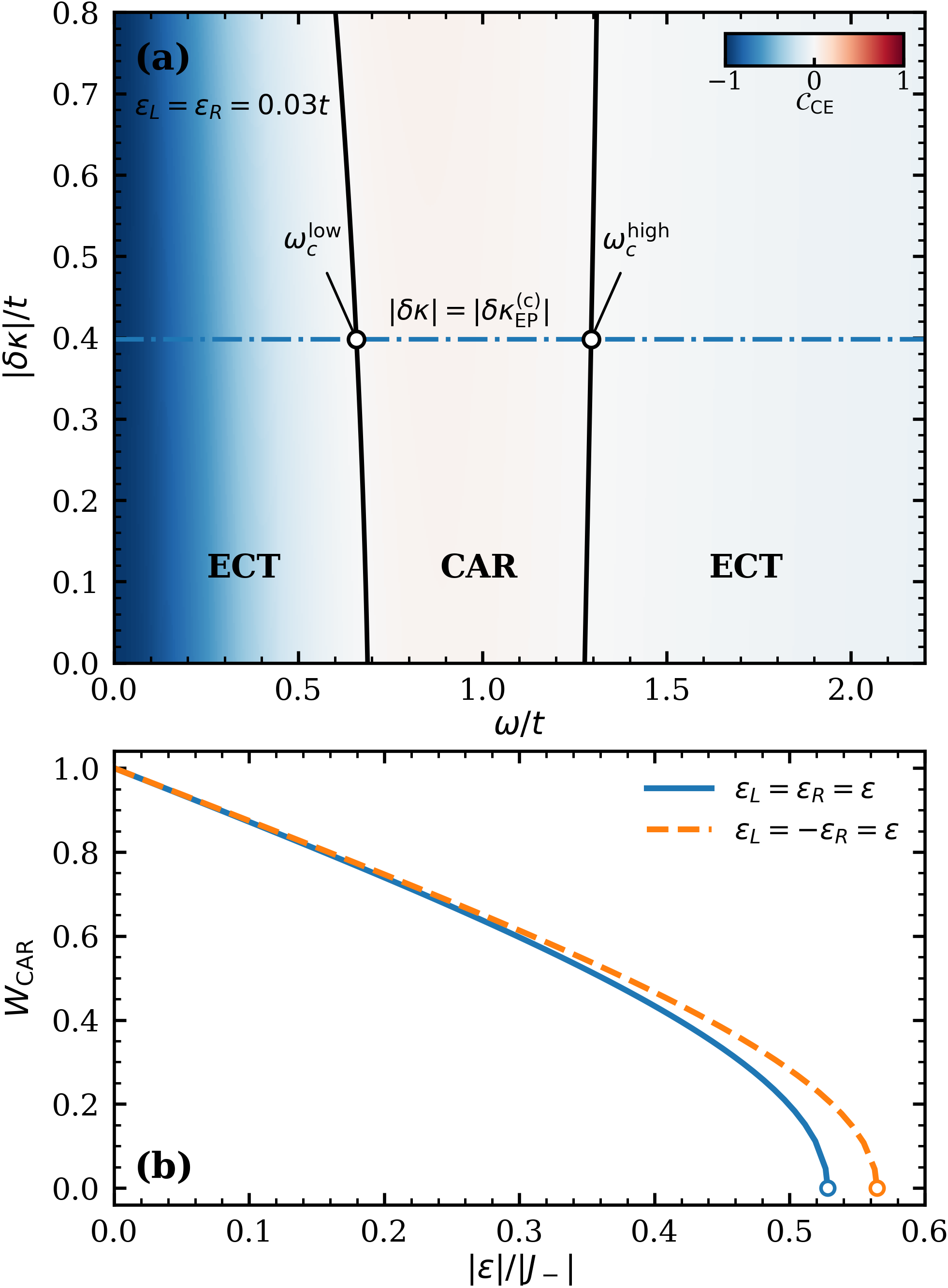}
\caption{
\textbf{Robustness of the Bogoliubov-interference structure to
onsite-energy detuning.} (a) CAR--ECT contrast $C_{\rm CE}$ evaluated from the full $4\times4$ retarded Green function for common detuning $\epsilon_L=\epsilon_R=0.03t$. Finite detuning mixes the two Bogoliubov sectors and removes the exact block decomposition. Solid curves mark $\mathcal{C}_{\rm CE}=0$, while the dash-dotted line denotes the shifted exceptional-point condition. A finite CAR-dominated region persists despite sector mixing.
(b) Normalized CAR-window width \( W_{\rm CAR} = (\omega_c^{\rm high}-\omega_c^{\rm low})/([\omega_c^{\rm high}-\omega_c^{\rm low}]_{\epsilon=0})\) evaluated along the corresponding shifted exceptional-point loci for common detuning $\epsilon_L=\epsilon_R=\epsilon$ and antisymmetric detuning $\epsilon_L=-\epsilon_R=\epsilon$. The horizontal axis is normalized by the weak-sector scale $J_-=t-\Delta$. In both cases the normalized CAR-window width decreases continuously rather than disappearing under infinitesimal detuning, demonstrating that the interference structure persists after the exact sector decomposition is broken. Parameters are $t=1$, $\Delta=0.9t$, and $\bar\kappa=t$.
}
\label{fig:SM-detuning}
\end{figure}

Away from resonance, the physical ECT and CAR amplitudes are evaluated
directly from the full Green function
\begin{equation}
G^R(\omega)
=
\left(
\omega-H_{\rm eff}^R
\right)^{-1},
\label{eq:SM-full-Green-detuned}
\end{equation}
without invoking the resonant identities $G_{ee}^{LR}=(G_++G_-)/2$ and $G_{eh}^{LR}=(G_+-G_-)/2$.  
We use \(T_{\rm ECT} = \Gamma_L^N\Gamma_R^N \left|G_{ee}^{LR}\right|^2\) and \(T_{\rm CAR} = \Gamma_L^N\Gamma_R^N \left|G_{eh}^{LR}\right|^2\), and determine the interference boundaries numerically from \(T_{\rm CAR}=T_{\rm ECT}\). 

Figure~\ref{fig:SM-detuning} shows that the ECT--CAR--ECT structure survives finite onsite-energy detuning even though the two Bogoliubov sectors are no longer exactly decoupled.  Along the shifted exceptional-point conditions, the CAR-dominated window remains finite under moderate sector mixing and decreases continuously with increasing onsite-energy detuning for both representative detuning directions. 
The exact $H_+\oplus H_-$ decomposition therefore provides a transparent analytic limit, while Fig.~\ref{fig:SM-detuning} demonstrates that the interference reorganization is not an artifact of exact sector conservation.

For the parameters used in Fig.~\ref{fig:SM-detuning}, $t=1$, $\Delta=0.9t$, and $\bar\kappa=t$, the natural weak-sector spectral scale is \(J_-=t-\Delta=0.1t \).

\subsection{Robustness of the synchronized intersection to sector-mixing detuning}
\label{sec:SM-synchronized-detuning}

The robustness analysis above shows that the ECT--CAR--ECT reorganization survives after onsite-energy detuning breaks the exact $H_+\oplus H_-$ sector decomposition. We now ask the stronger question whether the synchronized intersection between the spectral exceptional point and the interference bifurcation also survives.

For finite detuning we work directly with the full $4\times4$ retarded Green function, \(G^R(\omega)=\left[\omega-H_{\rm eff}^R\right]^{-1}\), 
and define
\begin{equation}
\Phi(\omega;\bar\kappa,d,\epsilon)
=
T_{\rm CAR}(\omega)-T_{\rm ECT}(\omega),
\qquad d\equiv|\delta\kappa|.
\end{equation}
A CAR--ECT interference bifurcation is characterized, without reference to an exact sector decomposition, by a double root,
\begin{equation}
\Phi(\omega_c;\bar\kappa_c,d_c,\epsilon)=0,
\qquad
\partial_\omega\Phi(\omega_c;\bar\kappa_c,d_c,\epsilon)=0.
\label{eq:SM-detuned-double-root}
\end{equation}

We constrain $d_c$ to the exceptional-point branch continuously connected to the resonant $H_-$ exceptional point. 
For common detuning, $\epsilon_L=\epsilon_R=\epsilon$, solving \(Q_-^{(c)}=0\) gives \(a^2=(\sqrt{t^2-\epsilon^2}\pm \Delta)^2\). Selecting the branch continuously connected to \(|a|=|t-\Delta|\) at \(\epsilon=0\) yields
\begin{equation}
d_c^{(c)}(\epsilon) = 4\left|\sqrt{t^2-\epsilon^2}-\Delta\right|.
\end{equation}
For antisymmetric detuning, $\epsilon_L=-\epsilon_R=\epsilon$, solving \(Q_-^{(a)}=0\) gives \(a^2=(t\pm \sqrt{\Delta^2-\epsilon^2})^2\). Selecting the branch continuously connected to \(|a|=|t-\Delta|\) at \(\epsilon=0\) yields
\begin{equation}
d_c^{(a)}(\epsilon) = 4\left|t-\sqrt{\Delta^2-\epsilon^2}\right|.
\end{equation}
At each detuning we then solve Eq.~(\ref{eq:SM-detuned-double-root}) for $\bar\kappa_c(\epsilon)$ and $\omega_c(\epsilon)$ using the full $4\times4$ Green function.

At resonance this procedure returns exactly
\begin{equation}
\bar\kappa_c(0)=\sqrt{2t\Delta},
\qquad
d_c(0)=4|t-\Delta|,
\qquad
\omega_c(0)=\sqrt{\frac{3t\Delta}{2}}.
\end{equation}
For weak detuning the synchronized solution evolves continuously away from this analytic point. For example, for common detuning $\epsilon/t=0.03$ and $t=1$, $\Delta=0.9t$, we obtain
\begin{equation}
\frac{\bar\kappa_c}{t}\simeq 1.133063,
\qquad
\frac{d_c}{t}\simeq 0.398200,
\qquad
\frac{\omega_c}{t}\simeq 0.991634.
\end{equation}
The two CAR--ECT boundaries coalesce at this shifted spectral EP and split again upon increasing $d$ through $d_c$, demonstrating that the $0\to1\to2$ interference-boundary bifurcation persists after exact sector conservation is broken.

The local persistence of the synchronized solution also follows from the implicit-function theorem. At the resonant synchronized point, the Jacobian of the two double-root conditions with respect to $(\bar\kappa,\omega)$ is non-singular.
For the parameters \(t=1\) and \(\Delta=0.9 t\) used above, numerical evaluation at the resonant synchronized point gives
\begin{equation}
\det
\frac{\partial\!\left(\Phi,\partial_\omega\Phi\right)}
{\partial\!\left(\bar\kappa,\omega\right)}
\bigg|_{\epsilon=0}
\simeq 5.881 \times 10^{-3}\neq0.
\label{eq:SM-detuned-IFT}
\end{equation}
Since the shifted exceptional-point branch is smooth for weak detuning, Eq.~(\ref{eq:SM-detuned-IFT}) guarantees a unique nearby synchronized solution for sufficiently small $|\epsilon|$.

\begin{figure}[t]
\centering
\includegraphics[width=0.42\linewidth]{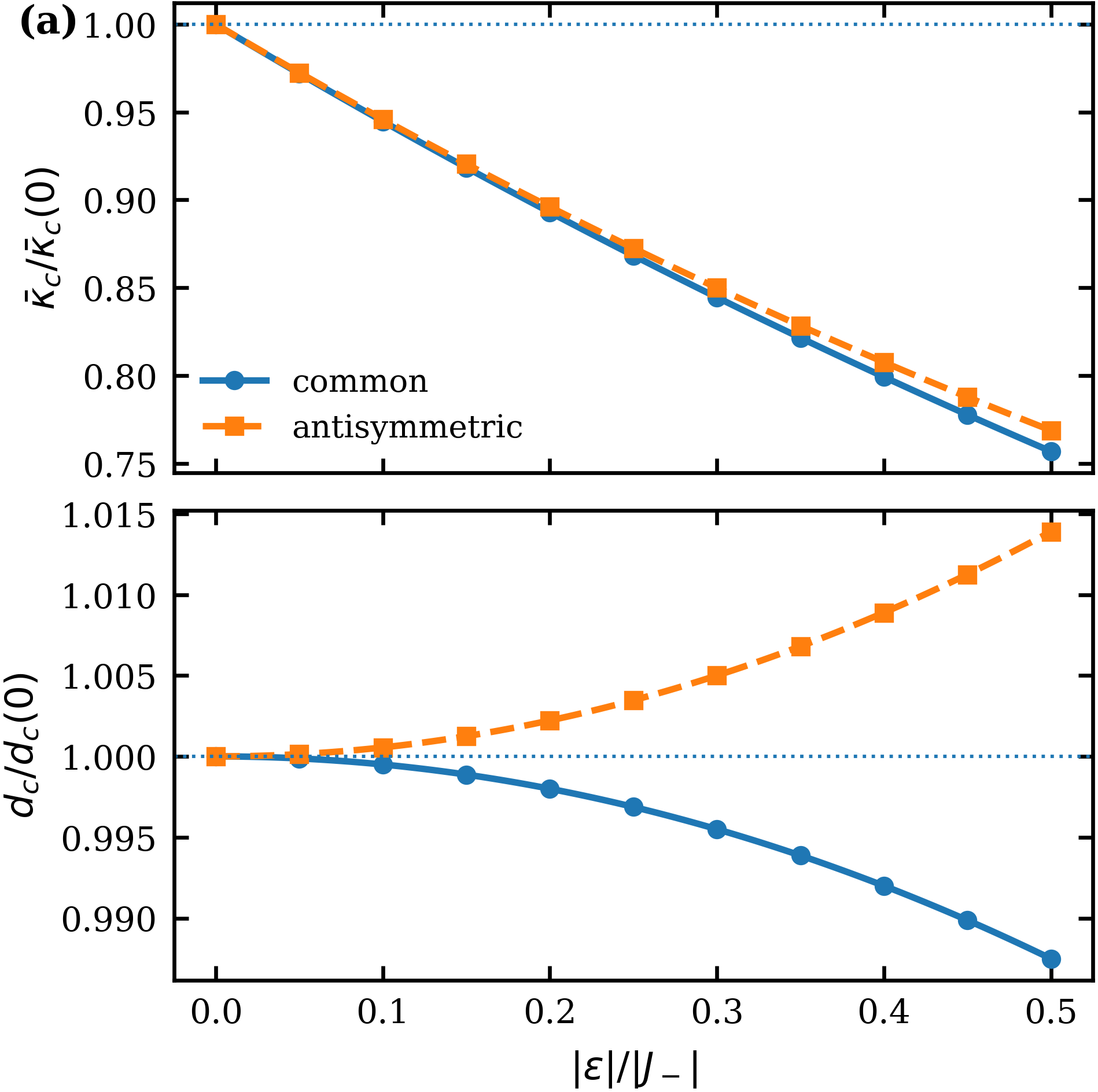} \,\,\,
\includegraphics[width=0.42\linewidth]{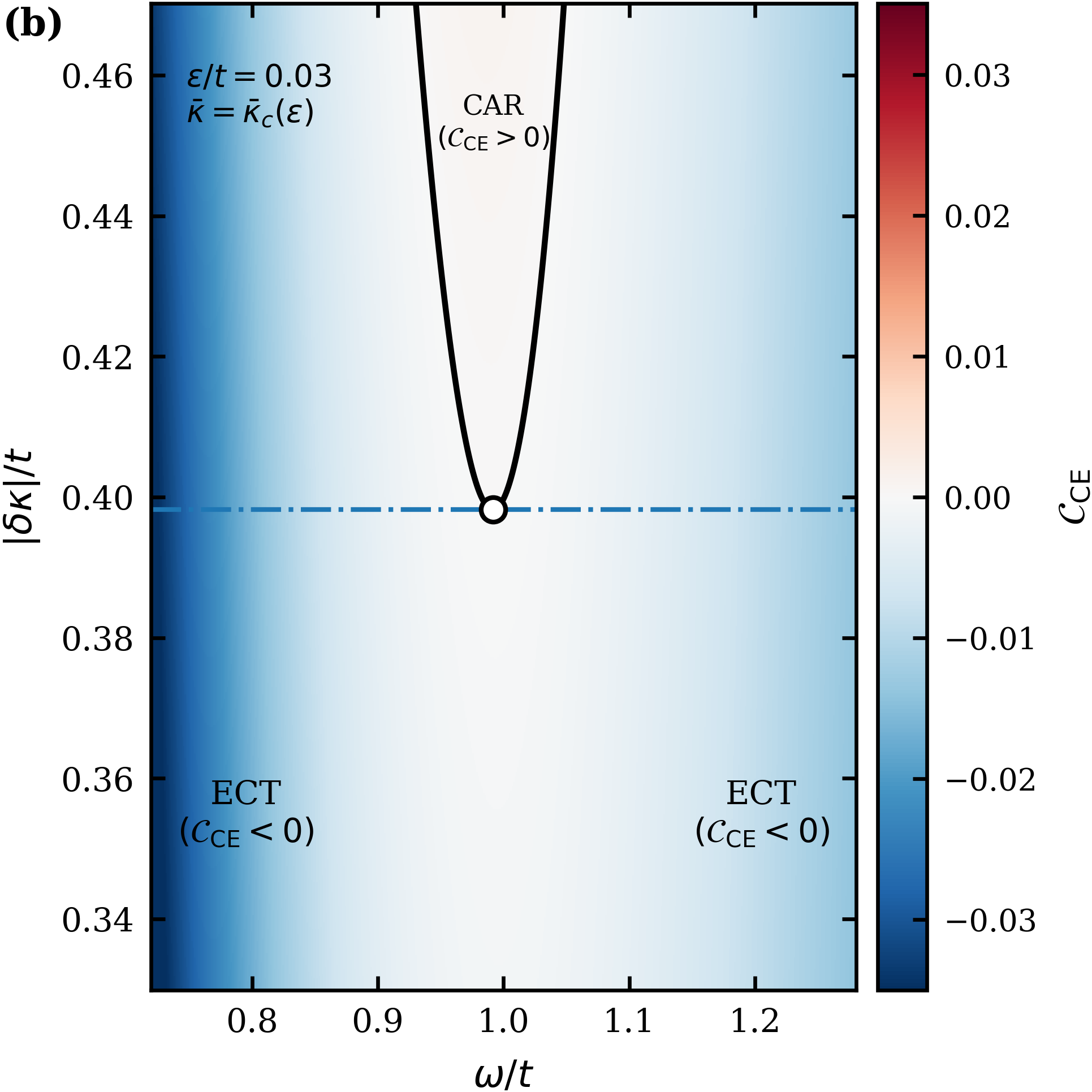}
\caption{Robustness of the synchronized spectral--interference
intersection to onsite-energy detuning.
(a) Continuation of the synchronized mean linewidth $\bar\kappa_c$ and spectral-EP imbalance $d_c$, normalized by their resonant values, for common and antisymmetric detuning. The detuning axis is normalized by the weak-sector scale $|J_-|=|t-\Delta|$.
(b) CAR--ECT contrast for representative common detuning $\epsilon/t=0.03$, evaluated at the retuned $\bar\kappa=\bar\kappa_c(\epsilon)$. The solid contour denotes ${\cal C}_{\rm CE}=0$, the dash-dotted horizontal line is the shifted spectral-EP condition $d=d_c(\epsilon)$, and the open circle marks their synchronized intersection. Parameters are $t=1$ and $\Delta=0.9t$.}
\label{fig:SM-sync-detuning}
\end{figure}

Figure~\ref{fig:SM-sync-detuning} summarizes this continuation. The synchronized spectral--interference intersection is therefore continuously displaced, rather than destroyed, by weak onsite-energy detuning; the mean linewidth $\bar\kappa$ provides the control required to restore exact synchronization once the resonant sector symmetry is broken.

\section{Transport observables and current correlations}
\label{sec:SM-transport}

This section connects the energy-resolved Bogoliubov-interference structure derived in Sec.~\ref{sec:SM-interference} to directly measurable transport observables. We first derive the nonlocal differential conductance and its finite-temperature convolution, then obtain the critical thermal scale associated with the synchronized interference bifurcation. Finally, we formulate the zero-frequency current cross correlations in the full BdG scattering framework and show that, for the lead-induced realization used in Fig.~3 of the main text, every CAR--ECT balance point is also a zero of the differential cross correlation.

\subsection{Nonlocal current and differential conductance}
\label{sec:SM-nonlocal-conductance}

The electron and hole distribution functions in normal lead
$\alpha=L,R$ are
\begin{equation}
f_\alpha^e(\omega)
=
f(\omega-eV_\alpha),
\qquad
f_\alpha^h(\omega)
=
f(\omega+eV_\alpha),
\label{eq:SM-fermi-eh}
\end{equation}
where the superconducting condensate defines the zero of energy. 

For the nonlocal processes relevant here, an electron incident from lead $R$ can be transmitted to lead $L$ either as an electron (elastic cotunneling, ECT) or as a hole (crossed Andreev reflection, CAR). With the current convention used in the main text, the nonlocal contribution can be written in Landauer--B\"uttiker form as~\cite{AnantramDatta96prb}
\begin{align}
I_L^{\rm NL}
=
\frac{e}{h}\int_0^\infty d\omega\,
\Big\{
&T_{\rm CAR}(\omega)
\left[
f_R^e(\omega)-f_L^h(\omega)
\right]
-
T_{\rm ECT}(\omega)
\left[
f_R^e(\omega)-f_L^e(\omega)
\right]
\Big\}
+\mathcal I_h ,
\label{eq:SM-nonlocal-current-general}
\end{align}
where $\mathcal I_h$ denotes the particle-hole-conjugate contribution obtained by exchanging electron and hole occupations and the corresponding BdG transmission channels. Equation~(\ref{eq:SM-nonlocal-current-general}) makes explicit the opposite charge signs with which CAR and ECT contribute to the nonlocal current.

For the one-sided bias protocol used in the main text, \(V_L=0\) and \(V_R=V\). In the positive-energy BdG representation of Eq.~(\ref{eq:SM-nonlocal-current-general}), the electron-incident contribution is proportional to \(\mathcal T_{\rm NL}^{e}(\omega) [f(\omega-eV)-f(\omega)]\), while the particle-hole-conjugate term is proportional to \(-\mathcal T_{\rm NL}^{h}(\omega) [f(\omega+eV)-f(\omega)]\), where \(\mathcal T_{\rm NL}^{e(h)}(\omega)\) denotes the CAR-minus-ECT transmission combination for electron (hole) incidence. BdG particle-hole symmetry gives \(\mathcal T_{\rm NL}^{h}(\omega) =\mathcal T_{\rm NL}^{e}(-\omega)\). Using \(f(-x)=1-f(x)\) and changing \(\omega\rightarrow-\omega\) in the hole contribution combines the two terms into the full-energy representation
\begin{align}
I_L^{\rm NL}
=
\frac{e}{h}\int_{-\infty}^{\infty} d\omega\,
\mathcal T_{\rm NL}(\omega)
[f(\omega-eV) - f(\omega)] ,
\label{eq:SM-nonlocal-current}
\end{align}
where \(\mathcal T_{\rm NL}(\omega) = T_{\rm CAR}(\omega)-T_{\rm ECT}(\omega)\). 
Within the voltage-independent scattering description used here, the bias enters only through the lead occupations. 
Differentiating with respect to \(V\), using \(\partial_V f(\omega-eV) =-e\,\partial_\omega f(\omega-eV)\), gives
\begin{equation}
G_{\rm NL}(V,T)
\equiv
\frac{dI_L^{\rm NL}}{dV}
=
\frac{e^2}{h}
\int_{-\infty}^{\infty} d\omega\,
\mathcal T_{\rm NL}(\omega)
\left[
-\partial_\omega f(\omega-eV)
\right] .
\label{eq:SM-GNL}
\end{equation}
Thus, positive nonlocal differential conductance corresponds to CAR-dominated transport after thermal averaging in the convention adopted here; at $T=0$, this reduces directly to the energy-resolved CAR--ECT imbalance. 

At zero temperature, \(-\partial_\omega f(\omega-eV) \rightarrow \delta(\omega-eV)\), 
so that
\begin{equation}
G_{\rm NL}(V,0)
=
\frac{e^2}{h}
\left[
T_{\rm CAR}(eV)-T_{\rm ECT}(eV)
\right].
\label{eq:SM-GNL-zeroT}
\end{equation}
The conductance zeros therefore coincide exactly with the energy-resolved CAR--ECT interference boundaries at $T=0$.

\subsection{Finite-temperature conductance}
\label{sec:SM-finite-temperature}

At finite temperature, the measured nonlocal conductance is a thermal convolution of the energy-resolved interference signal,
\begin{equation}
G_{\rm NL}(V,T)
=
\frac{e^2}{h}
\int d\omega\,
\mathcal T_{\rm NL}(\omega)
K_T(\omega-eV),
\label{eq:SM-thermal-convolution}
\end{equation}
with
\begin{equation}
K_T(\epsilon)
=
-\partial_\epsilon f(\epsilon)
=
\frac{1}{4k_BT}
\operatorname{sech}^2
\left(
\frac{\epsilon}{2k_BT}
\right).
\label{eq:SM-thermal-kernel}
\end{equation}

The experimentally resolved conductance boundaries are defined by
\begin{equation}
G_{\rm NL}
\left(
V_c^{\rm low/high}(T),T
\right)
=
0.
\label{eq:SM-finiteT-boundaries}
\end{equation}
At zero temperature they reduce to
\begin{equation}
eV_c^{\rm low/high}(0)
=
\omega_c^{\rm low/high},
\label{eq:SM-zeroT-boundaries}
\end{equation}
where $\omega_c^{\rm low/high}$ are the analytic interference boundaries derived in Sec.~\ref{sec:SM-interference-boundaries}. For $T>0$, however, the conductance zeros are properties of the thermally averaged response and need not coincide with the bare energy-resolved roots.

Near the synchronized interference bifurcation, the positive CAR-dominated interval becomes parametrically narrow as the exceptional point is approached from above. Thermal convolution therefore eventually removes the positive-conductance interval altogether. This defines a characteristic temperature $T^*$, whose near-critical behavior provides a directly measurable thermal signature of the same bifurcation.

\subsection{Critical thermal scale near the synchronized bifurcation}
\label{sec:thermal-critical-scaling}

We now derive the near-critical temperature scale associated with the EP-anchored interference bifurcation discussed in the main text. Throughout this subsection we impose the synchronized condition
\begin{equation}
\bar\kappa
=
\bar\kappa_c,
\qquad
\bar\kappa_c^2
=
\frac{J_+^2-J_-^2}{2}
=
2t\Delta ,
\label{eq:SM-kappac}
\end{equation}
and define \(d\equiv|\delta\kappa|\) and \(d_{\rm EP} \equiv |\delta\kappa|_{\rm EP} = 4|J_-|\). 

For the lead-induced realization, the zero-temperature nonlocal transmission imbalance can be written as
\begin{equation}
\mathcal T_{\rm NL}(\omega)
=
-\Gamma_L^N\Gamma_R^N
\frac{
J_+J_-\,\mathcal F_{\rm int}(\omega)
}{
|D_+(\omega)D_-(\omega)|^2
},
\label{eq:SM-TNL-factorized}
\end{equation}
where \(\mathcal F_{\rm int}(\omega) = (\omega^2-\Lambda_+)(\omega^2-\Lambda_-) + \bar\kappa^2\omega^2\). 
The denominator in Eq.~(\ref{eq:SM-TNL-factorized}) is smooth and nonzero in the vicinity of the interference bifurcation. Therefore, the local sign and critical structure of $\mathcal T_{\rm NL}$ are governed by $\mathcal F_{\rm int}$.

Writing \(x=\omega^2\), the interference polynomial is a monic quadratic in $x$ and can be expressed exactly as
\begin{equation}
\mathcal F_{\rm int}(x)
=
\left[
x-x_0(d)
\right]^2
-
\frac{\mathcal D_{\rm int}(d)}{4},
\label{eq:SM-F-normal-form}
\end{equation}
where $x_0(d)$ denotes the position of the quadratic minimum. At the synchronized linewidth,
\begin{equation}
\mathcal D_{\rm int}(d)
=
\frac{J_+^2-J_-^2}{8}
\left(
d^2-d_{\rm EP}^2
\right).
\label{eq:SM-Dint-critical}
\end{equation}
Hence $\mathcal D_{\rm int}<0$ below the EP, vanishes at the EP, and becomes positive above it.

Using the synchronized bifurcation energy derived in Sec.~\ref{sec:SM-critical-unfolding},
\begin{equation}
x_0(d_{\rm EP})
=
\omega_*^2
=
\frac{3}{8}
\left(
J_+^2-J_-^2
\right)
=
\frac{3}{2}t\Delta .
\label{eq:SM-omega-star}
\end{equation}
For $d>d_{\rm EP}$ sufficiently close to the EP, let \(\delta d = d-d_{\rm EP}>0\). 
Equation~(\ref{eq:SM-Dint-critical}) then gives
\begin{equation}
\mathcal D_{\rm int}
=
\left(
J_+^2-J_-^2
\right)
|J_-|\,\delta d
+
O(\delta d^2).
\label{eq:SM-Dint-expansion}
\end{equation}

To obtain the thermal scale, we expand the transmission imbalance near the center of the emerging CAR window. Denoting \(\omega_0(d)=\sqrt{x_0(d)}\) and writing \(\omega=\omega_0+\epsilon\), one has, to leading order,
\begin{equation}
x-x_0
=
\omega^2-\omega_0^2
=
2\omega_0\epsilon
+
O(\epsilon^2).
\label{eq:SM-x-local}
\end{equation}
Since the remaining prefactor in Eq.~(\ref{eq:SM-TNL-factorized}) is smooth, the local transmission profile takes the normal form
\begin{equation}
\mathcal T_{\rm NL}(\omega)
\simeq
\mathcal C_*
\left[
\frac{\mathcal D_{\rm int}}{4}
-
4\omega_*^2\epsilon^2
\right],
\qquad
\mathcal C_*>0,
\label{eq:SM-TNL-local-normal}
\end{equation}
up to higher-order corrections in $\delta d$ and $\epsilon$. Equation~(\ref{eq:SM-TNL-local-normal}) explicitly shows that the positive CAR-dominated response is born from a parabolic touching at the synchronized EP.

At finite temperature, the measured response follows from Eq.~(\ref{eq:SM-thermal-convolution}). Near the bifurcation, the maximum of the thermally broadened conductance occurs at \(eV=\omega_0+O(\delta d)\). 
Using the normalization and second moment of the thermal kernel 
\begin{align}
\int d\epsilon\,K_T(\epsilon)
&=1,
\\
\int d\epsilon\,\epsilon^2K_T(\epsilon)
&=
\frac{\pi^2}{3}(k_BT)^2,
\label{eq:SM-fermi-second-moment}
\end{align}
Eq.~(\ref{eq:SM-TNL-local-normal}) gives
\begin{equation}
G_{\rm NL}^{\rm max}(T)
\simeq
\frac{e^2}{h}\mathcal C_*
\left[
\frac{\mathcal D_{\rm int}}{4}
-
\frac{4\pi^2}{3}
\omega_*^2(k_BT)^2
\right].
\label{eq:SM-Gmax-local}
\end{equation}

We define $T^*$ as the temperature at which the positive-conductance interval disappears, equivalently \(G_{\rm NL}^{\rm max}(T^*)=0\). 
It follows that
\begin{equation}
(k_BT^*)^2
\simeq
\frac{
3\,\mathcal D_{\rm int}
}{
16\pi^2\omega_*^2
}.
\label{eq:SM-Tstar-Dint}
\end{equation}
Substituting Eqs.~(\ref{eq:SM-Dint-expansion}) and (\ref{eq:SM-omega-star}) yields
\begin{equation}
k_BT^*
\simeq
\sqrt{
\frac{|J_-|}{2\pi^2}
\left(
|\delta\kappa|-|\delta\kappa|_{\rm EP}
\right)
}
\label{eq:SM-Tstar-scaling}
\end{equation}
for
$|\delta\kappa|-|\delta\kappa|_{\rm EP}\rightarrow0^+$.
Thus
\begin{equation}
T^*
\propto
\left(
|\delta\kappa|-|\delta\kappa|_{\rm EP}
\right)^{1/2}.
\label{eq:SM-Tstar-exponent}
\end{equation}

This square-root law is the finite-temperature counterpart of the zero-temperature opening of the CAR-dominated energy window,
\begin{equation}
\Delta\omega_{\rm CAR}
\propto
\left(
|\delta\kappa|-|\delta\kappa|_{\rm EP}
\right)^{1/2}.
\label{eq:SM-window-scaling-reference}
\end{equation}
The two quantities therefore inherit the same critical exponent from the synchronized interference bifurcation. The numerical finite-temperature convolution shown in Fig.~3(b) of the main text gives \(\beta_{T,{\rm fit}}\simeq0.500\), in agreement with the analytic prediction $\beta_T=1/2$.

\subsection{BdG scattering formulation of zero-frequency current correlations}
\label{sec:SM-noise-formalism}

We next formulate the current fluctuations using the full
Bogoliubov--de Gennes scattering matrix in the channel basis
$(Le,Lh,Re,Rh)$~\cite{AnantramDatta96prb}. The retarded Green function of the central system is \(G^R(\omega) = \left[\omega-H_{\rm eff}^R\right]^{-1}\). 
Only the normal transport leads constitute scattering channels. Accordingly, their coupling matrix is
\begin{equation}
W
=
\operatorname{diag}
\left(
\sqrt{\Gamma_L^N},
\sqrt{\Gamma_L^N},
\sqrt{\Gamma_R^N},
\sqrt{\Gamma_R^N}
\right),
\label{eq:SM-W}
\end{equation}
whereas auxiliary loss enters $G^R$ through
$\kappa_\alpha=\Gamma_\alpha^N+\gamma_\alpha$ but not through $W$.

In the wide-band limit the scattering matrix is
\begin{equation}
S(\omega)
=
\mathbb I
-iWG^R(\omega)W .
\label{eq:SM-scattering-matrix}
\end{equation}
Charge currents in the two normal leads are represented by
\begin{equation}
Q_L
=
\operatorname{diag}(1,-1,0,0),
\qquad
Q_R
=
\operatorname{diag}(0,0,1,-1),
\label{eq:SM-charge-matrices}
\end{equation}
and we define
\begin{equation}
A_\alpha(\omega)
=
Q_\alpha
-
S^\dagger(\omega)Q_\alpha S(\omega).
\label{eq:SM-Aalpha}
\end{equation} 
For the incoming-channel occupation matrix \(\mathsf F(\omega)
=\operatorname{diag}
\left(
f_L^e,
f_L^h,
f_R^e,
f_R^h
\right)\), 
the zero-frequency symmetrized cross correlation is
\begin{equation}
S_{LR}
=
\frac{e^2}{h}
\int d\omega\,
\operatorname{Re}
\operatorname{Tr}
\left[
A_L\mathsf F A_R(\mathbb I-\mathsf F)
+
A_R\mathsf F A_L(\mathbb I-\mathsf F)
\right].
\label{eq:SM-SLR-general}
\end{equation}

For the one-sided zero-temperature bias protocol used in Fig.~3(c),
\begin{equation}
V_L=0,
\qquad
V_R=V>0,
\qquad
T=0,
\label{eq:SM-noise-bias}
\end{equation}
the only additional occupied incoming channel in the transport window
$0<\omega<eV$ is the right electron channel. The bias-dependent part
of the cross correlation may therefore be written as
\begin{equation}
S_{LR}(V)-S_{LR}(0)
=
\frac{e^2}{h}
\int_0^{eV}
d\omega\,
\mathcal S_{LR}(\omega),
\label{eq:SM-SLR-window}
\end{equation}
where $\mathcal S_{LR}(\omega)$ is the corresponding energy-resolved
cross-noise density. Consequently,
\begin{equation}
\frac{dS_{LR}}{d(eV)}
=
\frac{e^2}{h}
\mathcal S_{LR}(eV).
\label{eq:SM-dSLR}
\end{equation}

\subsection{Exact shared zeros of the differential cross correlation}
\label{sec:SM-noise-factorization}

We now establish analytically the relation between the current cross correlations and the CAR--ECT interference boundaries. In this subsection we restrict to the lead-induced realization used in Fig.~3(c) of the main text, \(\gamma_L=\gamma_R=0\) and \(\kappa_\alpha=\Gamma_\alpha^N\), for which the scattering problem is closed within the two normal leads and the BdG scattering matrix is unitary.

To distinguish the transport interference polynomial from the incoming occupation matrix, we denote the former by
\begin{equation}
\mathcal F_{\rm int}(\omega)
=
(\omega^2-\Lambda_+)(\omega^2-\Lambda_-)
+
\bar\Gamma^{\,2}\omega^2 ,
\label{eq:SM-Fint}
\end{equation}
where \(\Lambda_s = J_s^2 + \frac{\Gamma_L^N\Gamma_R^N}{4}\) and \(\bar\Gamma = \frac{\Gamma_L^N+\Gamma_R^N}{2}\). 
The energy-resolved CAR--ECT imbalance is then
\begin{equation}
T_{\rm CAR}(\omega)-T_{\rm ECT}(\omega)
=
-\Gamma_L^N\Gamma_R^N
\frac{
J_+J_-\,\mathcal F_{\rm int}(\omega)
}{
|D_+(\omega)D_-(\omega)|^2
}.
\label{eq:SM-Tdifference-Fint}
\end{equation}

Substituting the scattering matrix of Eq.~(\ref{eq:SM-scattering-matrix}) into the full BdG cross-correlation formula Eq.~(\ref{eq:SM-SLR-general}), and evaluating the trace for the one-sided zero-temperature occupation matrix, gives an exact rational function of energy,
\begin{equation}
\mathcal S_{LR}(\omega)
=
\frac{
\mathcal P_{\rm noise}(\omega)
}{
\mathcal D_{\rm noise}(\omega)
}.
\label{eq:SM-noise-rational}
\end{equation}
For arbitrary $J_\pm$, $\Gamma_L^N$, and $\Gamma_R^N$ within this lead-induced model, its numerator factorizes as 
\begin{equation}
\mathcal P_{\rm noise}(\omega) =
\mathcal F_{\rm int}(\omega)
\mathcal N_{\rm noise}(\omega).
\label{eq:SM-noise-numerator-factorization}
\end{equation}
Equivalently,
\begin{equation}
\mathcal S_{LR}(\omega)
=
\mathcal F_{\rm int}(\omega)
\mathcal R_{\rm noise}(\omega),
\label{eq:SM-noise-exact-factorization}
\end{equation}
where \(\mathcal R_{\rm noise}(\omega)
\equiv \mathcal N_{\rm noise}(\omega)/\mathcal D_{\rm noise}(\omega)\). 
For the finite lead broadenings considered here, \(\mathcal D_{\rm noise}(\omega)\) is nonsingular at the real-energy interference roots. 
The factor $\mathcal R_{\rm noise}$ contains additional two-particle interference information specific to the current fluctuations. Its zeros can therefore generate differential-noise sign changes with no counterpart in the average CAR--ECT transport imbalance. 

For completeness, the origin of the additional noise zero shown in Fig.~3(c) of the main text can be made explicit.  At the parameters used in that panel,
\begin{equation}
t=1,\qquad
\Delta=0.9t,\qquad
\bar\Gamma=\bar\kappa_c=\sqrt{2t\Delta},
\qquad
|\delta\Gamma|=t ,
\label{eq:SM-noise-Fig3c-parameters}
\end{equation}
the numerator of the exact cross-noise density factorizes, up to an overall nonzero constant, as
\begin{equation}
\mathcal P_{\rm noise}(\omega)
\propto
\mathcal F_{\rm int}(\omega)\,
\mathcal N_{\rm noise}^{(3c)}(\omega).
\label{eq:SM-noise-Fig3c-factorization}
\end{equation}
Introducing the dimensionless frequency
\(u=\omega/t\), the two factors may be written as
\begin{align}
\mathcal F_{\rm int}(u)
&\propto
160000u^4-415200u^2+254241,
\label{eq:SM-noise-Fig3c-Fint}
\\
\mathcal N_{\rm noise}^{(3c)}(u)
&\propto
25600000000u^8
-132864000000u^6
+181927360000u^4
\nonumber\\
&\quad
+\left(
99532800000\sqrt{5}
-24746750400
\right)u^2
-49485213999 .
\label{eq:SM-noise-Fig3c-Nnoise}
\end{align}
The positive roots of the interference factor are
\begin{equation}
\frac{\omega_c^{\rm low}}{t}
\simeq0.995033,
\qquad
\frac{\omega_c^{\rm high}}{t}
\simeq1.266850,
\label{eq:SM-noise-Fig3c-shared-roots}
\end{equation}
which are precisely the two CAR--ECT balance points inherited by the differential cross correlation. In contrast, \(\mathcal N_{\rm noise}^{(3c)}\) has the additional positive root
\begin{equation}
\frac{\omega_n}{t}
\simeq0.462892 ,
\label{eq:SM-noise-Fig3c-extra-root}
\end{equation}
producing the noise-specific zero marked by the dash-dotted line in Fig.~3(c). This explicit specialization illustrates both directions of the factorization: the CAR--ECT boundaries are necessarily shared zeros, whereas the fluctuation factor can supply additional zeros that have no counterpart in the average transport imbalance.

Equation~(\ref{eq:SM-noise-exact-factorization}) immediately implies that every CAR--ECT balance point is a zero of the differential cross correlation under the conditions specified above:
\begin{equation}
\mathcal F_{\rm int}(\omega_c)=0
\quad\Longrightarrow\quad
\mathcal S_{LR}(\omega_c)=0.
\label{eq:SM-shared-noise-zero}
\end{equation}
Combining Eqs.~(\ref{eq:SM-GNL-zeroT}), (\ref{eq:SM-dSLR}), and (\ref{eq:SM-Tdifference-Fint}) therefore gives
\begin{equation}
G_{\rm NL}(V_c,0)=0
\quad\Longrightarrow\quad
\left.
\frac{dS_{LR}}{d(eV)}
\right|_{eV=\omega_c}
=0 .
\label{eq:SM-shared-measurable-zeros}
\end{equation}

The converse does not generally hold: additional roots of $\mathcal N_{\rm noise}(\omega)$ can produce zeros of the differential cross correlation without satisfying $T_{\rm CAR}=T_{\rm ECT}$. Current fluctuations therefore inherit the Bogoliubov-interference boundaries governing the average nonlocal transport while retaining additional fluctuation-specific structure.

For finite auxiliary loss, the normal-lead scattering submatrix is not a closed unitary scattering problem. A complete fluctuation treatment must then include the corresponding dissipative bath channels and their occupations. The exact factorization Eq.~(\ref{eq:SM-noise-exact-factorization}) and the shared-zero statement Eq.~(\ref{eq:SM-shared-measurable-zeros}) are therefore stated here specifically for the lead-induced realization and one-sided zero-temperature protocol used in Fig.~3(c) of the main text.

\section{Experimental implementation and parameter requirements}
\label{sec:SM-experimental}

The effective model considered here maps naturally onto hybrid semiconductor--superconductor quantum-dot devices.  The coherent normal coupling $t$ and nonlocal pairing amplitude $\Delta$ correspond, respectively, to effective elastic-cotunneling and crossed-Andreev-reflection couplings between neighboring dots.
Both processes, including electrostatic control of their relative strength, have been demonstrated experimentally in hybrid nanowire devices~\cite{BordinKouwenhovenDvir23prx,BordinDvir24prl}. The normal-lead rates $\Gamma_{L,R}^{N}$ are controlled by the corresponding dot--lead tunnel barriers, while the auxiliary rates $\gamma_{L,R}$ in Eq.~(\ref{eq:SM-total-linewidth}) represent additional local dissipative channels.

\subsection{Lead-induced realization}

If no auxiliary loss is used, $\gamma_L=\gamma_R=0$, so that
$\kappa_\alpha=\Gamma_\alpha^{N}$. 
The synchronized operating conditions derived in Sec.~\ref{sec:SM-critical-unfolding}, \(\bar\kappa_c=\sqrt{2t\Delta}\) and \(|\delta\kappa|_{\rm EP}=4|t-\Delta|\), 
then translate directly into
\begin{equation}
\bar\Gamma^{N}=\sqrt{2t\Delta},
\qquad
|\Gamma_L^{N}-\Gamma_R^{N}|
=4|t-\Delta|.
\label{eq:SM-exp-lead-condition}
\end{equation}
Equivalently,
\begin{equation}
\Gamma_{L,R}^{N}
=
\sqrt{2t\Delta}
\pm2|t-\Delta|,
\label{eq:SM-exp-lead-rates}
\end{equation}
up to interchange of $L$ and $R$.
For the representative parameters used in the main text,
$\Delta=0.9t$, this gives
\begin{equation}
\frac{\Gamma_L^{N}}{t}\simeq1.542,
\qquad
\frac{\Gamma_R^{N}}{t}\simeq1.142,
\qquad
\frac{\Gamma_L^{N}}{\Gamma_R^{N}}\simeq1.35 .
\label{eq:SM-exp-lead-numbers}
\end{equation}
Thus the required left--right asymmetry is moderate, although the
absolute lead broadening is comparable to the coherent coupling
scale.

\subsection{Auxiliary-loss realization}

A more flexible implementation separates transport injection and readout from spectral linewidth control.  Let the two normal-lead couplings be symmetric, i.e., \(\Gamma_L^{N}=\Gamma_R^{N}\equiv\Gamma^{N}\), and introduce independently tunable local loss rates $\gamma_L$ and $\gamma_R$. 
Defining \(\bar\gamma=\frac{\gamma_L+\gamma_R}{2}\) and \(\delta\gamma=\gamma_L-\gamma_R\), 
one has
\begin{equation}
\bar\kappa=\Gamma^{N}+\bar\gamma,
\qquad
\delta\kappa=\delta\gamma .
\end{equation}
The synchronized conditions therefore become
\begin{equation}
\Gamma^{N}+\bar\gamma=\sqrt{2t\Delta},
\qquad
|\delta\gamma|=4|t-\Delta| .
\label{eq:SM-exp-loss-condition}
\end{equation}
Because $\gamma_L$ and $\gamma_R$ are passive loss rates, they must remain nonnegative. Using $\gamma_{L,R}=\bar\gamma\pm\delta\gamma/2$, the synchronized operating point therefore requires
\begin{equation}
\bar\gamma\ge\frac{|\delta\gamma|_{\rm EP}}{2}
=2|t-\Delta|,
\end{equation}
or equivalently
\begin{equation}
\Gamma^N
\le
\sqrt{2t\Delta}-2|t-\Delta|.
\label{eq:SM-exp-loss-positivity}
\end{equation}
For $\Delta=0.9t$, this gives $\Gamma^N/t\le1.142$.  At the limiting value $\Gamma^N=\sqrt{2t\Delta}-2|t-\Delta|$, the synchronized point can be reached with a single auxiliary loss channel, $\gamma_L=4|t-\Delta|$ and $\gamma_R=0$ (or with $L$ and $R$ interchanged). 
This realization allows the normal leads to remain optimized for transport spectroscopy while the mean and differential loss are used as independent spectral-control parameters.  In particular, choosing
\begin{equation}
\gamma_L=\bar\gamma+\frac{\delta\gamma}{2},
\qquad
\gamma_R=\bar\gamma-\frac{\delta\gamma}{2}
\end{equation}
and scanning $\delta\gamma$ at fixed $\bar\gamma$ follows precisely the fixed-$\bar\kappa$ trajectory used in Fig.~2 of the main text.
Engineered dissipation and tunable coupling to environmental channels have been demonstrated in superconducting quantum circuits, including the experimental realization and control of dissipation-induced exceptional points~\cite{ChenMurch22prl,AbbasiMurch22prl}.
Applying this reservoir-engineering principle to hybrid quantum-dot transport provides a route to implementing the independent $\gamma_\alpha$ controls assumed here.

\subsection{Experimental protocol and signatures}

A possible tuning procedure is as follows. First, the coherent couplings are calibrated spectroscopically and tuned to $t\simeq\Delta$, producing the scale separation $|J_-|\ll|J_+|$ required for sector selectivity. The mean linewidth is then tuned across $\bar\kappa_c=\sqrt{2t\Delta}$, either through the normal-lead barriers or through the combination of transport and auxiliary-loss channels, while the differential linewidth is swept through the relevant spectral and interference thresholds. Below and above \(\bar{\kappa}_c\), the relative ordering of these two thresholds is reversed, providing a direct experimental signature of the control-space unfolding shown in Fig.~2(a) of the main text. At \(\bar{\kappa} = \bar{\kappa}_c\), synchronization requires the two signatures to occur at the same control point: the relevant spectral resonances coalesce at the exceptional point, while the two CAR--ECT balance boundaries simultaneously merge. On crossing this point, the latter split and produce the finite positive-nonlocal-conductance interval discussed in Fig.~3 of the main text. Combining spectral spectroscopy with nonlocal differential conductance therefore provides both a verification of controlled synchronization and a distinction from a generic CAR--ECT crossover.

\subsection{Energy scales and temperature}

Existing hybrid quantum-dot experiments establish coherent ECT/CAR couplings, their gate tunability, and local/nonlocal transport spectroscopy~\cite{BordinKouwenhovenDvir23prx,BordinDvir24prl}. 
As a representative experimental scale, a two-dot Kitaev-chain device has been quantitatively described by effective couplings \(t=\Delta\simeq 12\,\mu\)eV, normal-lead broadening \(\Gamma_L=\Gamma_R\simeq 4\,\mu\)eV, and an electron temperature of approximately \(45\) mK~\cite{DvirKouwenhoven23nature}. Using \(t=12\,\mu\)eV as a demonstrated coherent-coupling scale and retaining the representative ratio \(\Delta/t=0.9\) employed in the main text, the synchronized conditions correspond to
\begin{equation}
\bar{\kappa}_c\simeq 16.1\,\mu{\rm eV},
\qquad
|\delta\kappa|_{\rm EP}\simeq 4.8\,\mu{\rm eV}.
\end{equation}
For the purely lead-induced realization this gives
\begin{equation}
\Gamma_L^N\simeq 18.5\,\mu{\rm eV},
\qquad
\Gamma_R^N\simeq 13.7\,\mu{\rm eV},
\end{equation}
illustrating that the required left--right asymmetry is moderate, although stronger absolute lead coupling than in the above benchmark device is required. Alternatively, auxiliary loss permits the normal transport contacts to remain weaker while independently supplying the required mean and differential linewidths. 
Hybrid-dot implementations demonstrate effective ECT/CAR couplings on the scale of \(10\,\mu{\rm eV}\)~\cite{DvirKouwenhoven23nature}, while strongly proximitized YSR-based devices have reached excitation gaps above \(70\,\mu{\rm eV}\)~\cite{ZatelliDvir24ncommun}, indicating that larger coherent energy scales are experimentally accessible.

The temperature requirement is more restrictive close to the synchronized bifurcation. From Eq.~(\ref{eq:SM-Tstar-scaling}), 
\begin{equation}
k_BT^*
\simeq
\sqrt{
\frac{|J_-|}{2\pi^2}
\left(
|\delta\kappa|-|\delta\kappa|_{\rm EP}
\right)
}, \notag
\end{equation}
and therefore \(T^*\) vanishes on approaching the critical point. Resolving the bifurcation itself and resolving its asymptotic square-root thermal regime are thus distinct experimental requirements: the latter requires sufficiently low electron temperature or a larger coherent coupling scale. At finite distance above threshold the CAR-dominated window broadens and becomes correspondingly more thermally accessible. Weak dot-level detuning does not impose an additional fine-tuning requirement, since Sec.~\ref{sec:SM-synchronized-detuning} and Fig.~\ref{fig:SM-sync-detuning} show that it continuously displaces rather than destroys the synchronized intersection; retuning \(\bar{\kappa}\) restores exact synchronization.

\section{Jordan-critical conditional dynamics}
\label{sec:SM-Jordan-dynamics}

We derive here the time-domain dynamics associated with the sector-selective exceptional point underlying the synchronized interference bifurcation. The central result is that an appropriately chosen cross-site Bogoliubov channel eliminates the zeroth-order identity contribution to the propagator and directly isolates the Jordan term at the EP. 
The derivation below also separates this intrinsic Jordan dynamics from the common passive-decay envelope and clarifies the roles of sector-selective preparation, conditional normalization, and postselection survival.

\subsection{Exact sector propagator}

In the Bogoliubov-sector basis, the relevant block can be written as
\begin{equation}
H_-
=
-\frac{i\bar\kappa}{2}I+K_-,
\qquad
K_-
=
J_-\tau_x
-i\frac{\delta\kappa}{4}\tau_z .
\label{eq:SM-Hminus-dynamics}
\end{equation}
The traceless part satisfies
\begin{equation}
K_-^2
=
\left(
J_-^2-\frac{\delta\kappa^2}{16}
\right)I
\equiv
\Omega_-^2 I .
\label{eq:SM-Ksquare}
\end{equation}
This identity permits the propagator to be evaluated exactly.

For \(|\delta\kappa|<4|J_-|\), the quantity
\begin{equation}
\Omega_-
=
\sqrt{
J_-^2-\frac{\delta\kappa^2}{16}
}
\label{eq:SM-Omega}
\end{equation}
is real. Since \(K_-^2=\Omega_-^2 I\), separating the even and odd powers of \(K_-\) in the exponential series gives
\begin{equation}
U_-(\tau)
=
e^{-iH_-\tau}
=
e^{-\bar\kappa\tau/2}
\left[
\cos(\Omega_-\tau)I
-i
\frac{\sin(\Omega_-\tau)}{\Omega_-}
K_-
\right].
\label{eq:SM-U-underdamped}
\end{equation}

At the exceptional point, \(|\delta\kappa|=4|J_-|\), one has \(K_-^2=0\). 
For $J_-\neq0$, however, $K_-\neq0$ and $\operatorname{rank}K_-=1$ at $|\delta\kappa|=4|J_-|$.
Hence the defective sector is similar to a single $2\times2$ Jordan block, $J_2(-i\bar\kappa/2)$. 
The exponential series therefore terminates exactly,
\begin{equation}
U_-^{\rm EP}(\tau)
=
e^{-\bar\kappa\tau/2}
\left(
I-i\tau K_-
\right).
\label{eq:SM-U-EP}
\end{equation}
The term linear in \(\tau\) is the time-domain manifestation of the nontrivial Jordan block at the EP2.

For \(|\delta\kappa|>4|J_-|\), we define
\begin{equation}
\chi_-
=
\sqrt{
\frac{\delta\kappa^2}{16}-J_-^2
}. 
\label{eq:SM-chi}
\end{equation}
Analytic continuation of Eq.~(\ref{eq:SM-U-underdamped}) then gives
\begin{equation}
U_-(\tau)
=
e^{-\bar\kappa\tau/2}
\left[
\cosh(\chi_-\tau)I
-i
\frac{\sinh(\chi_-\tau)}{\chi_-}
K_-
\right].
\label{eq:SM-U-overdamped}
\end{equation}

Equations~(\ref{eq:SM-U-underdamped})--(\ref{eq:SM-U-overdamped}) therefore exhibit the continuous evolution of the propagator kernel,
\begin{equation}
\frac{\sin(\Omega_-\tau)}{\Omega_-}
\longrightarrow
\tau
\longrightarrow
\frac{\sinh(\chi_-\tau)}{\chi_-}
\label{eq:SM-kernel-crossover}
\end{equation}
from oscillatory dynamics through the Jordan-critical point to
overdamped evolution.

\subsection{Jordan-selective transition channel}

Let \(|-_{\rm N}\rangle = (|e\rangle-|h\rangle)/\sqrt{2}\) denote the Bogoliubov state selecting the \(H_-\) sector. We consider the preparation and readout
\begin{equation}
|i\rangle=|R,-_{\rm N}\rangle,
\qquad
|f\rangle=|L,-_{\rm N}\rangle.
\label{eq:SM-Jordan-channel}
\end{equation}
In the basis
\(\{|L,-_{\rm N}\rangle,|R,-_{\rm N}\rangle\}\),
\begin{equation}
K_-
=
\begin{pmatrix}
-i\delta\kappa/4 & J_-\\
J_- & i\delta\kappa/4
\end{pmatrix}.
\label{eq:SM-Kmatrix}
\end{equation}
The selected channel obeys
\begin{equation}
\langle f|i\rangle=0,
\qquad
\langle f|K_-|i\rangle=J_-.
\label{eq:SM-channel-properties}
\end{equation}
Consequently, at the exceptional point the identity contribution in Eq.~(\ref{eq:SM-U-EP}) vanishes exactly, leaving
\begin{equation}
{\cal A}_{L\leftarrow R}^{\rm EP}(\tau)
=
\langle f|U_-^{\rm EP}(\tau)|i\rangle
=
-iJ_-\tau\,e^{-\bar\kappa\tau/2}.
\label{eq:SM-EP-amplitude}
\end{equation}
The corresponding unnormalized transition probability is therefore
\begin{equation}
P_{L,\rm EP}^{\rm raw}(\tau)
=
J_-^2\tau^2 e^{-\bar\kappa\tau}.
\label{eq:SM-EP-raw}
\end{equation}
Thus the cross-site sector-selective channel eliminates the
zeroth-order identity contribution exactly and converts the
nilpotent Jordan term into the leading transition amplitude at the EP.

\subsection{Raw, survival, and conditional dynamics}

For definiteness, we take \(J_->0\) and \(\delta\kappa>0\); the other sign choices follow analogously. Starting from \(|R,-_{\rm N}\rangle\), the state in the oscillatory regime is
\begin{equation}
|\psi(\tau)\rangle
=
e^{-\bar\kappa\tau/2}
\begin{pmatrix}
-iJ_-\dfrac{\sin(\Omega_-\tau)}{\Omega_-}
\\[2mm]
\cos(\Omega_-\tau)
+
\dfrac{\delta\kappa}{4}
\dfrac{\sin(\Omega_-\tau)}{\Omega_-}
\end{pmatrix}.
\label{eq:SM-state-under}
\end{equation}
The raw left-site probability is therefore
\begin{equation}
P_L^{\rm raw}(\tau)
=
e^{-\bar\kappa\tau}
J_-^2
\frac{\sin^2(\Omega_-\tau)}{\Omega_-^2},
\label{eq:SM-PLraw-under}
\end{equation}
while the survival probability is
\begin{align}
P_{\rm surv}(\tau)
=
\langle\psi(\tau)|\psi(\tau)\rangle
=
e^{-\bar\kappa\tau}
\left\{
J_-^2
\frac{\sin^2(\Omega_-\tau)}{\Omega_-^2}
+
\left[
\cos(\Omega_-\tau)
+
\frac{\delta\kappa}{4}
\frac{\sin(\Omega_-\tau)}{\Omega_-}
\right]^2
\right\}.
\label{eq:SM-survival-under}
\end{align}
Conditioning on survival gives
\begin{equation}
P_L^{(c)}(\tau)
=
\frac{
J_-^2
\dfrac{\sin^2(\Omega_-\tau)}{\Omega_-^2}
}{
J_-^2
\dfrac{\sin^2(\Omega_-\tau)}{\Omega_-^2}
+
\left[
\cos(\Omega_-\tau)
+
\dfrac{\delta\kappa}{4}
\dfrac{\sin(\Omega_-\tau)}{\Omega_-}
\right]^2
}.
\label{eq:SM-Pconditional-under}
\end{equation}

In the overdamped regime the corresponding expressions follow by the analytic continuation
\begin{equation}
\frac{\sin(\Omega_-\tau)}{\Omega_-}
\rightarrow
\frac{\sinh(\chi_-\tau)}{\chi_-},
\qquad
\cos(\Omega_-\tau)
\rightarrow
\cosh(\chi_-\tau).
\label{eq:SM-analytic-continuation}
\end{equation}

Exactly at the exceptional point, taking \(\delta\kappa=4J_-\), one obtains
\begin{equation}
|\psi_{\rm EP}(\tau)\rangle
=
e^{-\bar\kappa\tau/2}
\begin{pmatrix}
-iJ_-\tau\\
1+J_-\tau
\end{pmatrix},
\label{eq:SM-state-EP}
\end{equation}
and hence
\begin{equation}
P_{\rm surv}^{\rm EP}(\tau)
=
e^{-\bar\kappa\tau}
\left[
(J_-\tau)^2+(1+J_-\tau)^2
\right].
\label{eq:SM-survival-EP}
\end{equation}
The exact conditional transfer law becomes
\begin{equation}
P_{L,\rm EP}^{(c)}(\tau)
=
\frac{(J_-\tau)^2}
{(J_-\tau)^2+(1+J_-\tau)^2}.
\label{eq:SM-Pconditional-EP}
\end{equation}

The common factor \(e^{-\bar\kappa\tau}\) cancels upon conditioning, whereas the differential non-Hermitian term proportional to \(\delta\kappa\) remains in the normalized dynamics. 
Conditioning therefore removes only the common passive-decay envelope; it does not remove the differential non-Hermitian structure that generates the exceptional point and its Jordan dynamics.

\begin{figure}[t]
\centering
\includegraphics[width=0.4\columnwidth]{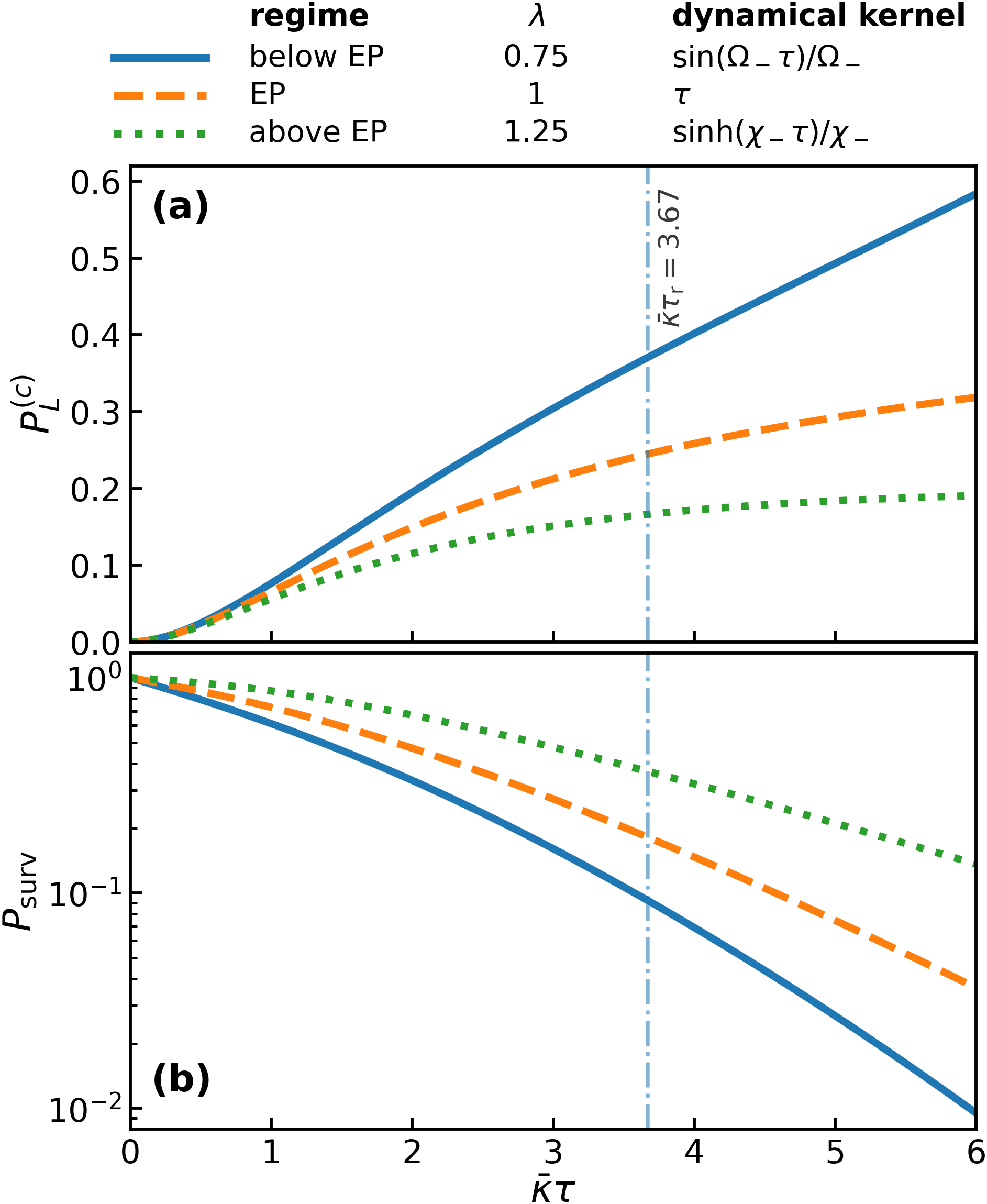} 
\caption{(Color online) {\bf Jordan-critical conditional dynamics.} (a) Conditional transfer probability \(P_L^{(c)}\) across the sector-selective \(H_-\) exceptional point. The three curves correspond to
\(\lambda\equiv \delta\kappa/\delta\kappa_{\rm EP}^{(-)} =0.75,1,1.25\), exhibiting oscillatory, Jordan-critical, and overdamped kernels, respectively, as summarized above the panel. (b) Corresponding no-jump survival probabilities. The vertical line marks the representative readout time \(\bar\kappa\tau_{\rm r}=3.67\), illustrating the tradeoff between conditional contrast and postselection survival. Here \(t=1\) and \(\Delta=0.9t\) are the same coherent parameters as in the transport analysis, while the smaller mean linewidth \(J_-/\bar\kappa=0.36\) is chosen to enhance conditional dynamical visibility. 
}
\label{fig:FigS4}
\end{figure}

Figure~\ref{fig:FigS4} shows this conditional crossover together with the corresponding no-jump survival probability. The resulting oscillatory–Jordan-critical–overdamped crossover provides an independent dynamical fingerprint of the same sector-selective exceptional point that controls the steady-state Bogoliubov interference.

\subsection{Role of Bogoliubov-sector selectivity}

A conventional electron state is not sector selective. Since
\begin{equation}
|R,e\rangle
=
\frac{
|R,+_{\rm N}\rangle+|R,-_{\rm N}\rangle
}{\sqrt{2}},
\label{eq:SM-electron-injection}
\end{equation}
and analogously for left-electron readout, the corresponding
transition amplitude contains both sectors,
\begin{equation}
{\cal A}_{ee}^{LR}(\tau)
=
\frac{1}{2}
\left[
{\cal A}_+^{LR}(\tau)
+
{\cal A}_-^{LR}(\tau)
\right].
\label{eq:SM-electron-amplitude}
\end{equation}

For the coherent parameters used in the main text, \(t=1\) and \(\Delta=0.9t\), 
\begin{equation}
J_-=0.1,
\qquad
J_+=1.9,
\qquad
\frac{J_+}{J_-}=19.
\label{eq:SM-sector-ratio}
\end{equation}
When the \(H_-\) sector is tuned through its exceptional point, the \(H_+\) sector therefore remains strongly off critical and contributes a rapidly oscillating coherent background to ordinary electron injection and readout.  Preparing and measuring \(|-_{\rm N}\rangle\) suppresses this background and isolates the critical sector.

\subsection{Conditional visibility, survival, and parameters of Fig.~\ref{fig:FigS4}}

The conditional Jordan signature must be considered together with the probability of successful postselection. A useful dimensionless control parameter is \(r=\frac{J_-}{\bar\kappa}\), which compares the intrinsic Jordan time scale $J_-^{-1}$ with the common passive-decay time scale $\bar\kappa^{-1}$. Increasing $r$ allows the critical-sector dynamics to develop before postselection becomes exponentially unlikely.

To make the Jordan-critical crossover visible before the common passive-decay envelope suppresses the postselected signal, Fig.~\ref{fig:FigS4} uses the same coherent parameters \(t=1\) and \(\Delta=0.9 t\) as the transport analysis but a smaller mean linewidth than the synchronized transport value \(\bar{\kappa}_c=\sqrt{2t\Delta}\). This does not shift the \(H_-\) exceptional point, whose condition \(|\delta\kappa|_{\rm EP}^{(-)}=4|J_-|\) is independent of \(\bar{\kappa}\). 
For Fig.~\ref{fig:FigS4} we use \(\frac{J_-}{\bar\kappa}=0.36\), and compare
\begin{equation}
\frac{\delta\kappa}
{\delta\kappa_{\rm EP}^{(-)}}
=
0.75,\quad1,\quad1.25,
\qquad
\delta\kappa_{\rm EP}^{(-)}=4J_-.
\label{eq:SM-Fig4-lambda}
\end{equation}
Together with \(t=1\) and \(\Delta=0.9t\), this corresponds to
\(\bar\kappa/t\simeq0.278\). 
At the representative readout time \(\bar\kappa\tau\simeq3.67\), the conditional transfer probabilities for the three cases are approximately
\begin{equation}
P_L^{(c)}
\simeq
0.370,\quad0.245,\quad0.166,
\label{eq:SM-Fig4-Pc-values}
\end{equation}
while the corresponding survival probabilities are
\begin{equation}
P_{\rm surv}
\simeq
0.093,\quad0.182,\quad0.368.
\label{eq:SM-Fig4-survival-values}
\end{equation}

The same parameter choice remains strongly sector selective. The exceptional-point imbalance of the \(H_+\) sector is \(\delta\kappa_{\rm EP}^{(+)} = 4J_+ = 7.6\,t\), whereas the largest imbalance used in Fig.~\ref{fig:FigS4} is only \(\delta\kappa=0.5\,t\). Thus the largest applied imbalance is only approximately \(6.6\%\) of the \(H_+\) exceptional-point scale, leaving the \(H_+\) sector far from criticality.

The dynamics displayed in Fig.~\ref{fig:FigS4} can therefore be attributed to the Jordan-critical $H_-$ sector without simultaneously approaching the exceptional point of the spectator $H_+$ sector.